\documentclass[11pt,a4paper]{article}

\usepackage[margin=1in]{geometry}
\usepackage[T1]{fontenc}
\usepackage[utf8]{inputenc}
\usepackage{lmodern}
\usepackage{microtype}
\usepackage{hyphenat}
\usepackage{graphicx}
\usepackage{booktabs}
\usepackage[font=small,labelfont=bf,labelsep=period]{caption}
\usepackage{array}
\usepackage{multirow}
\usepackage{amsmath,amssymb,bm}
\usepackage{algorithm}
\usepackage{algpseudocode}
\usepackage{float}
\usepackage[section]{placeins}
\usepackage[numbers,sort&compress]{natbib}
\usepackage[hidelinks]{hyperref}
\usepackage[nameinlink,noabbrev]{cleveref}
\usepackage{longtable}
\usepackage{pdflscape}

\newcommand\figalttext[2][]{\relax}

\title{Physiology-Informed Sparse Identification of Dynamic Crop Models}
\author{
Corn\'{e} Verburg\textsuperscript{1,*}\thanks{Corresponding author: \href{mailto:c.verburg@tudelft.nl}{c.verburg@tudelft.nl}} \and
Alexander Heinlein\textsuperscript{1} \and
Fred van Eeuwijk\textsuperscript{2} \and
Neil Budko\textsuperscript{1}\\[0.5em]
\textsuperscript{1}Delft Institute of Applied Mathematics, Delft University of Technology, The Netherlands\\
\textsuperscript{2}Applied Statistics, Wageningen University and Research, The Netherlands
}
\date{}

\begin{document}
\maketitle

\makeatletter
\let\arxivoriginallabel\label
\renewcommand{\label}[1]{\arxivoriginallabel{#1}\arxivoriginallabel{main-#1}}
\makeatother

\begin{abstract}
  Developing resilient crops requires models that accurately capture how plant
  growth dynamics emerge from the continuous interplay between genetic and
  environmental factors. While high-throughput phenotyping platforms generate
  data for dynamic growth models, current modelling approaches show limitations:
  process-based crop growth models (CGMs) capture environmental dynamics but are
  difficult to scale across many genotypes because they require calibration of
  many parameters, while quantitative genetic models allow modelling many
  genotypes but often summarise time-varying environments using static
  covariates. To address this gap, we propose a data-driven framework to identify
  reduced systems of ordinary differential equations (ODEs) describing crop
  growth dynamics as functions of environmental variables and a controllable
  number of genotype-specific coefficients (GSCs). Using the Sparse Identification
  of Nonlinear Dynamics (SINDy) method, we infer model structure directly from
  data via sparse regression on a physiologically informed library of candidate
  functions. In this framework, GSCs parameterise cultivar-specific differences
  in response to environmental conditions. The resulting models are linear in
  these GSCs, making them compatible with mixed-model and genomic prediction
  approaches. We evaluate the approach on an \textit{in silico} dataset generated
  with WOFOST for five potato cultivars across twelve growing seasons. Our
  identified ODEs generalise to held-out years and achieve $R^2 > 0.9$ for
  end-of-season yield prediction across held-out years when compared to full-scale
  WOFOST simulations. These results provide a proof of concept for SINDy-based
  identification of parsimonious dynamic crop models from longitudinal
  crop-growth data.
\end{abstract}

\noindent\textbf{Keywords:} Crop growth modelling; genotype-environment
interaction; system identification; sparse regression; ordinary differential
equations; hybrid crop models.

\section{Introduction}\label{sec:introduction}
The emergence of high-throughput phenotyping data, obtained from sources such
as drone imagery and environmental sensors, has created new opportunities to
study plant growth dynamics over time~\citep{chawade2019high,jangra2021high}.
These data provide detailed information on the temporal development of
phenotypic traits and can improve yield prediction across a wide range of
environmental conditions. However, translating such longitudinal data into
models that relate genetic and environmental variation to phenotypic
trajectories is
challenging~\citep{araus2018,stockle2020can,moreira2020integrating}. This
challenge is particularly relevant for plant breeding, where the goal is to
identify genotypes that maintain relatively high yield under environmental
variation, i.e., \textit{resilient
    genotypes}~\citep{cooper2023breeding,langridge2021breeding}.

For breeding purposes, models for longitudinal phenotypic traits should be
functions of genetic and environmental inputs. The environmental inputs are
dynamic themselves and differ between trials, while genetic inputs are fixed
for a given genotype and may ultimately be represented by molecular marker
variation. Traditionally, crop growth models (CGMs) have been used to predict
and analyse crop growth dynamics in varying
environments~\citep{chapman2008use}. Process-based, mechanistic CGMs, such as
WOFOST (WOrld FOod STudies)~\citep{van1989wofost}, APSIM (Agricultural
Production Systems sIMulator)~\citep{keating2003overview}, and GECROS
(Genotype-by-Environment interaction on CROp growth
Simulator)~\citep{yin2005crop}, describe crop responses to environmental and
management variables through intermediate phenotypic traits, such as leaf area
index (LAI), rooting depth, and storage organ weight
(WSO)~\citep{wallach2018working}, in order to predict target traits such as
yield. For breeding applications, where the focus is on predicting yield across
a wide range of (trial) environments and genotypes, the detailed mechanistic
structure of traditional CGMs often cannot be calibrated with available data.
This motivates a more compact dynamic representation with a limited number of
genotype-specific parameters (GSPs), so that genetic differences in responses
to environmental variables can be estimated from data.

Although process-based CGMs are valuable tools for describing crop responses to
environmental and management conditions, they are difficult to use in breeding
applications. Their large number of variables and parameters makes calibration
challenging: parameters are weakly identifiable, and the resulting estimation
problem is often ill-posed. As a result, calibration typically requires
extensive field trials for each new
genotype~\citep{spitters1989crop,wang2013parameter,he2017data,van2016simulation,ech2025advances,ma2023variability,velluet2024practical}.
These limitations hinder the use of CGMs in realistic large-scale breeding
applications~\citep{weih2022application}.

A further difficulty arises when modelling the genetic component of G$\times$E
interactions using process-based CGMs. In many such models, output traits are
nonlinear and often implicit functions of the GSPs, complicating their
integration with linear mixed-model approaches for genomic prediction. A
well-known approach that integrates an existing CGM framework with genomic
prediction is the CGM-WGP (crop growth model -- whole genome prediction)
described in~\citep{technow2015integrating,messina2018leveraging}
and~\citep{jighly2023using}. The CGM-WGP uses the structure of the CGM to model
the dynamic correlation structure between traits and environmental inputs and
defines a small number of yield components as GSPs that depend linearly on
marker variation that can be predicted by genomic prediction. Nonlinear
genotype-to-phenotype relations, G$\times$E and epistatic interactions all
emerge from the structure of the CGM. A related strategy to CGM-WGP is
functional
mapping~\citep{wu2003functional,wu2006functional,pandey2021functional}. Instead
of CGMs, this approach uses a priori known systems of differential equations to
model the dynamics of yield components. A slightly simpler approach decouples
the dynamic modelling of yield components and estimation of GSPs from the
prediction of GSPs by marker variation. Examples of this two-step approach are
given by~\citep{yin2005qtl,poudel2023integrating}. However, estimating GSPs by
fitting CGMs to trait data is not only computationally demanding, but also
often ill-posed: inverse calibration can fit observed phenotypes while
recovering substantially incorrect GSP values~\citep{acharya2017reliability},
and weakly identifiable GSPs may therefore be poor targets for genomic
prediction~\citep{poudel2026co}. Scaling this calibration to breeding
populations further leads to high-dimensional optimisation problems that do not
scale well to breeding trials with hundreds of genotypes and multiple
environments~\citep{lamsal2017efficient}.

An alternative approach is to use static linear mixed models for yield with
reaction norm formulations for
G$\times$E~\citep{jarquin2014reaction,millet2019genomic,de2020data,piepho2024factor}.
In these models, differential genotypic sensitivity to environmental variables
of G$\times$E is at the basis of genotypic differences in adaptability and
resilience~\citep{van2016should}. For this approach to be successful, suitable
environmental variables need to be identified. Time-varying environmental
conditions are then aggregated into static variables~\citep{chapman2008use},
typically resulting in a loss of information about the dynamic nature of crop
growth. Other quantitative-genetic approaches, such as longitudinal and
random-regression models, can represent temporal or environmental variation
explicitly~\citep{jarquin2014reaction,millet2019genomic,de2020data,piepho2024factor},
but generally do not formulate crop development as an explicit dynamical system
of states driven by time-varying environmental variables.

Purely data-driven approaches, such as random forests and neural networks,
provide another alternative by modelling phenotypes directly as functions of
environmental variables~\citep{van2020crop,zhang2023machine}. While these
methods can capture complex nonlinear relationships, they generally do not
yield a small set of interpretable parameters that can be linked to genomic
information~\citep{hu2024climate}. Additional challenges include spatial
heterogeneity in field trials, limited availability of high-quality training
data, difficulty incorporating process knowledge, and poor extrapolation to
unseen
environments~\citep{lischeid2022machine,zhang2023machine,droutsas2022integration,hu2024climate,hu2023crop}.

Another related direction combines mechanistic crop models with machine
learning through hybrid or knowledge-guided
approaches~\citep{zheng2025mechanistic,li2026beyond}. In these approaches,
physiological knowledge can be incorporated using strategies such as CGM-based
input features, model architecture, or constraints in the loss
function~\citep{li2026beyond}. Depending on the formulation, such approaches
often retain either the parameter calibration complexity and computational
burden of the underlying CGM or the large number of parameters associated with
machine-learning models~\citep{zheng2025mechanistic}.

A promising direction, highlighted in recent
work~\citep{zhang2023machine,morales2023using,shahhosseini2021coupling}, is the
development of parsimonious dynamic models that combine a compact mathematical
structure with data-driven parameter identification. In this work, we test
SINDy as a framework for identifying reduced crop-growth ODEs directly from
longitudinal crop and environmental data. The framework is designed to produce
models that:
\begin{enumerate}
    \item describe crop growth dynamics under time-varying environmental variables,
    \item allow explicit control of model complexity,
    \item are linear in their parameters, facilitating future integration with genomic
          prediction methods,
    \item are amenable to stochastic extensions and statistical analysis.
\end{enumerate}

To meet these requirements, we consider a dynamic model that differs in form
from traditional process-based CGMs. At the level of aggregate traits such as
LAI and WSO, it may not be necessary to explicitly model all underlying
physiological processes. Instead, these processes can be approximated using
simpler terms that capture the most important interactions of growth with
environmental variables. Indeed, several studies have shown that simpler
empirical models can achieve comparable or even superior predictive performance
at certain
scales~\citep{soltani2015comparison,estes2013comparing,palosuo2011simulation,hajirad2025crop,kallenberg2023integrating,tardieu2017plant,hammer2006models}.

\begin{figure}[H]
    \centering
    \includegraphics[width=0.95\linewidth]{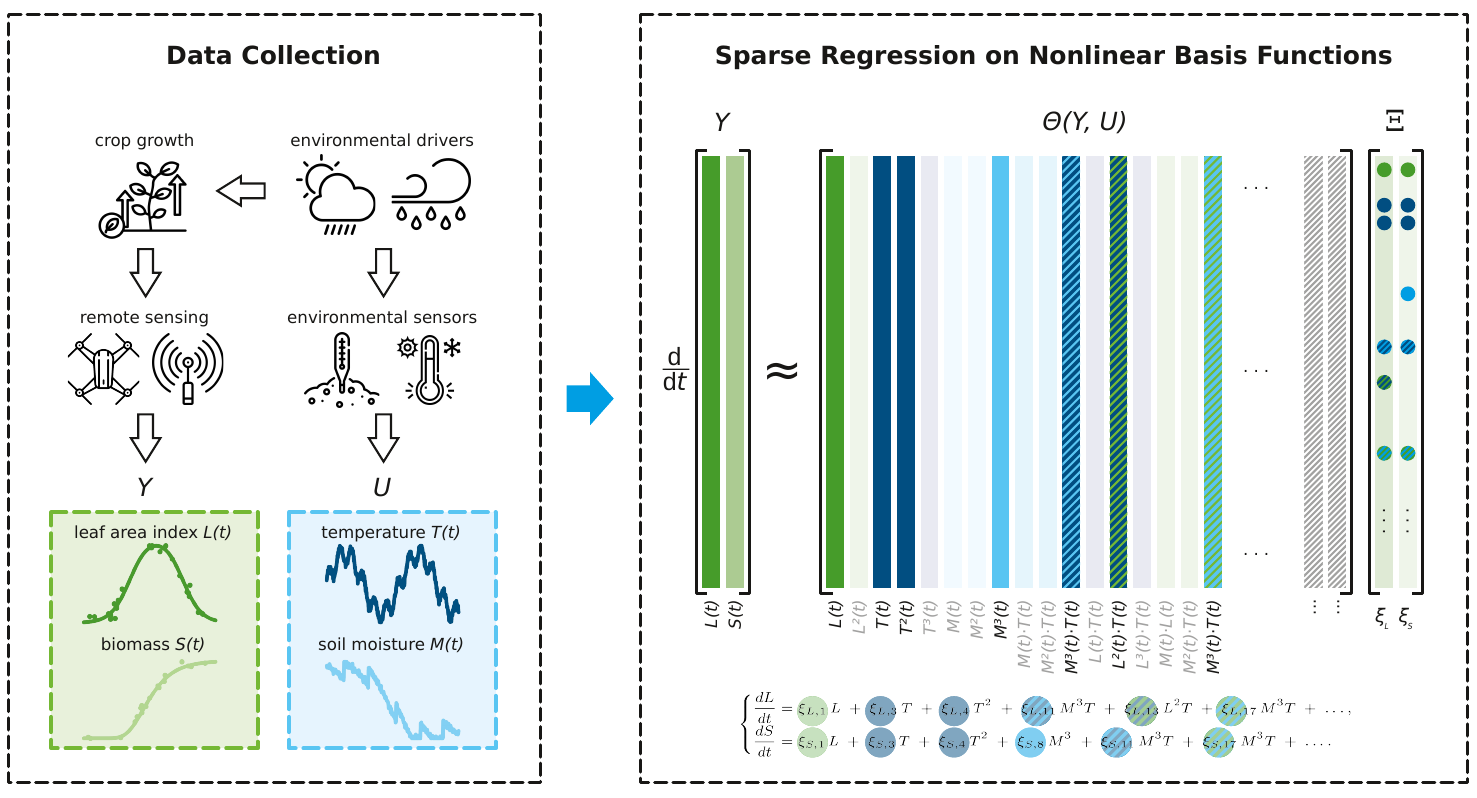}
    \caption{Workflow of the proposed SINDy-based approach. Data sources, such as remote sensing and environmental sensors, provide time series of crop traits, including leaf area index $L(t)$ and storage-organ weight $S(t)$, and environmental variables, such as temperature $T(t)$ and soil moisture $M(t)$. Sparse regression on nonlinear, user-selected basis functions collected in the library matrix $\Theta(Y, U)$ is used to estimate the coefficient matrix $\Xi$ and identify ordinary differential equations describing crop-trait dynamics.}\label{fig:pipeline_visualization}
    \figalttext[
        Workflow from crop and environmental time-series data to sparse dynamic crop models.
    ]{
        Workflow diagram showing crop and environmental time-series data being transformed into a library of nonlinear candidate functions, followed by sparse regression to identify differential equations for crop growth.
    }
\end{figure}

One approach to constructing such models is to exploit the smooth temporal
dynamics of traits such as LAI and WSO, which can often be approximated using
functions with relatively few parameters~\citep{prasad1993polynomial}. For
example, in~\citep{van2023conceptual}, the authors show that crop growth can be
described using a modified logistic curve with a small number of parameters
incorporating environmental effects. This suggests that crop growth can be
represented by a system of ODEs, in which the rate of change of a trait depends
on the trait itself and the environmental variables.

From a mathematical perspective, this interpretation is supported by the
standard state-rate formulation of dynamic crop growth models, in which state
variables are updated from process rates depending on the current state and
environmental variables~\citep{wallach2018working}. In this formulation, the
coefficients of the ODE system parameterise genotype-specific responses to
environmental variables.

Several studies have explored ODE-based crop growth models directly. For
example, in \citep{pelak2017dynamical}, the authors proposed a small system of
ODEs derived from the AquaCrop model~\citep{steduto2009aquacrop}, and other
parsimonious ODE-based models have been introduced
in~\citep{poudel2024parsimonious,zhao2019simple}. A key advantage of ODE-based
models is their continuous-time, explicit structure, which enables analytical
tools such as sensitivity analysis, identifiability analysis, and stability
analysis under environmental
variation~\citep{kuijpers2019model,pelak2017dynamical}. Moreover, the evolution
of trait distributions can be described through partial differential equations
derived from the underlying
ODEs~\citep{li2016probability,weisse2010quantifying}.

However, in existing work, the structure of the ODEs is typically specified
based on expert knowledge, following a \textit{hypothesis-driven} approach. The
systematic, data-driven identification of parsimonious ODE-based models has not
yet been widely explored in crop growth modelling. In this work, we address
this gap by using a \textit{data-driven} approach in which the structure of the
ODEs is identified directly from data using sparse regression techniques. This
enables the systematic exploration of a large space of candidate models and the
selection of parsimonious models with controllable complexity that capture the
essential dynamics of crop growth under varying environmental conditions.

We apply the Sparse Identification of Nonlinear Dynamics (SINDy)
framework~\citep{brunton2016discovering,messenger2021weak,mangan2017inferring,de2020pysindy,fasel2021sindy,wu2025data}
to identify systems of ODEs from longitudinal data. As a controlled proof of
concept, we evaluate the approach on an \textit{in silico} dataset generated
using the WOFOST model for potato. This setting allows us to assess predictive
performance, model sparsity, and stability under known environmental variation.

A schematic overview of the proposed approach is shown in
Figure~\ref{fig:pipeline_visualization}. Given time-series data of crop traits
and environmental variables, SINDy constructs a library of candidate functions
and uses sparse regression to identify a system of ODEs with controlled
complexity. The resulting model retains a reduced subset of environmental
variables, state variables, and their interactions that is selected to describe
the observed dynamics. We use the term \textit{physiology-informed} for our
approach to indicate that physiological knowledge guides the choice of state
variables, environmental drivers, and candidate-function structure.
Furthermore, the approach is \textit{data-driven} in the sense that the active
terms within the predefined library are identified directly from data rather
than specified a priori based on expert knowledge.

The resulting models are linear in their parameters, which makes them
compatible with mixed-model and genomic prediction approaches for
genotype-specific coefficients, an integration we leave to future work within
the CropXR program.

\section{Material and Methods}\label{sec:methods}
\subsection{Crop growth as a system of ODEs}
In this paper, as a proof of concept, we evaluate our ODE-identification
framework on growth curves generated \textit{in silico} by the well-established
process-based crop model WOFOST (WOrld FOod
STudies)~\citep{van1989wofost,de201925}. We focus on potato. Although
performance beyond this CGM remains to be evaluated, the formulation itself is
not specific to WOFOST or potato and can be applied to other traits, crops, and
CGMs.

For a fixed genotype, most process-based CGMs, including WOFOST, can be written
as discrete-time nonlinear state-space models of the form:
\begin{align}
    \label{eq:WOFOSTstep}
    \bm{y}_{i+1} = \text{CGM}(\bm{y}_i, \bm{u}_i; \bm{\kappa}), \qquad \bm{y}(t_p) = \bm{y}_0,
\end{align}
where $\bm{y}_i \in \mathbb{R}^n$ is the vector of state variables at time step
$t_i$, $\bm{u}_i \in \mathbb{R}^q$ is the (known) vector of environmental and
management (driving) variables, $\bm{\kappa} \in \mathbb{R}^s$ is the vector of
crop- and genotype-specific parameters (GSPs), and $\text{CGM}:\mathbb{R}^n
    \times \mathbb{R}^q \times \mathbb{R}^s \to \mathbb{R}^n$ is a nonlinear
function describing the one-step crop dynamics. The initial state at planting
time $t_p$ is given by $\bm{y}_0$. The state vector $\bm{y}_i$ includes
physiological traits such as organ biomass and developmental indicators, as
well as intermediate variables (e.g., assimilation rates) that are typically
not directly observable.

In the continuous-time limit, WOFOST can be viewed as the initial-value problem
\begin{align}
    \frac{d\bm y}{dt} = \bm R(\bm y(t), \bm u(t), \bm \kappa), \qquad \bm y(t_p)=\bm y_0.
    \label{eq:WOFOST-ODE}
\end{align}
However, the full state vector $\bm{y}$ is high-dimensional and contains many
internal variables that are not directly observable. In this work, we instead
focus on a small number of aggregate, high-level traits. For such traits, it is
plausible that the combined effect of lower-level physiological processes can
be captured implicitly, leading to a reduced-order model with fewer variables
and parameters, while still capturing the main dynamics and their response to
environmental variables.

Specifically, we consider two observable traits: LAI, denoted by $L$, and WSO,
denoted by $S$. These traits are selected due to their direct relation to light
interception and yield, respectively. Moreover, LAI is particularly relevant
for high-throughput phenotyping, as it is related to canopy structure and can
be estimated from drone imagery, which enables low-cost data collection for
model calibration.

The functions $S(t)$ and $L(t)$ are assumed to satisfy the following system of
ODEs:
\begin{align}
    \label{eq:SINDy-ODE-system}
    \begin{split}
        \frac{d S}{dt} & = R_S(\bm y, \bm u, \bm\kappa), \qquad S(t_0) = S_0, \\
        \frac{d L}{dt} & = R_L(\bm y, \bm u, \bm\kappa), \qquad L(t_0) = L_0,
    \end{split}
\end{align}
where $R_S$ and $R_L$ are the rate functions for WSO and LAI, respectively,
depending on the state variables $\bm{y}$, the driving variables $\bm{u}$, and
the GSPs $\bm{\kappa}$. The initial conditions $S_0$ and $L_0$ correspond to
the values of $S(t)$ and $L(t)$ at the start time of the simulation, $t_0$.

Our goal is to identify data-driven expressions for $R_S$ and $R_L$ that are
substantially simpler than their counterparts in WOFOST, while still accurately
describing the observed dynamics of $S(t)$ and $L(t)$. The SINDy framework
imposes the following, linear-in-parameters model structure:
\begin{align}
    \label{eq:LinComb_rate}
    R_{S}(\bm y, \bm u) & = \sum_{j=1}^{\ell}\xi_{j,S}\theta_{j}(t) + e_{S}(t), \\
    R_{L}(\bm y, \bm u) & = \sum_{j=1}^{\ell}\xi_{j,L}\theta_{j}(t) + e_{L}(t),
\end{align}
where $\theta_{j}(t)$ are candidate functions from a predefined library,
$\xi_{j,S}$ and $\xi_{j,L}$ are the corresponding coefficients, and $e_{S}(t)$
and $e_{L}(t)$ denote model approximation errors. The functions $\theta_{j}(t)$
depend on the state and driving variables, i.e., $\theta_{j}(t) =
    \theta_{j}(\bm y(t), \bm u(t))$. In this formulation, environmental effects are
captured through these nonlinear functions, while genotype-specific variation
is represented by the genotype-specific coefficients (GSCs) $\xi_{j,S}$ and
$\xi_{j,L}$. Details on the construction of the function library are provided
in Section~\ref{subsubsec:library_construction}, and further details on the
SINDy algorithm are given in the supplementary material. We refer to this SINDy
framework as \textit{physiology-informed} because physiological knowledge
guides the choice of state variables, environmental drivers, and
candidate-function structure, while the selection of active terms and
estimation of their coefficients remain \textit{data-driven}.

\subsection{Generating data with WOFOST}\label{subsec:data_sets}
To identify the rate functions $R_S$ and $R_L$, we require time series of the
state variables $S(t)$ and $L(t)$, together with environmental variables
$\bm{u}(t)$. We generate an \textit{in silico} dataset using WOFOST~8.1, as
implemented in the \texttt{pcse} library~\citep{de2018pcse,van1989wofost},
parameterised for five potato cultivars following~\citep{ten2022modelling}:
Festien, Fontane, Innovator, Markies, and Premiere. Simulations were performed
under water-limited production conditions without nitrogen limitation (using
the \texttt{Wofost81\_WLP\_CWB} model configuration as implemented in
\texttt{pcse 6.0.13}).

\begin{figure}[H]
    \includegraphics[width=\linewidth]{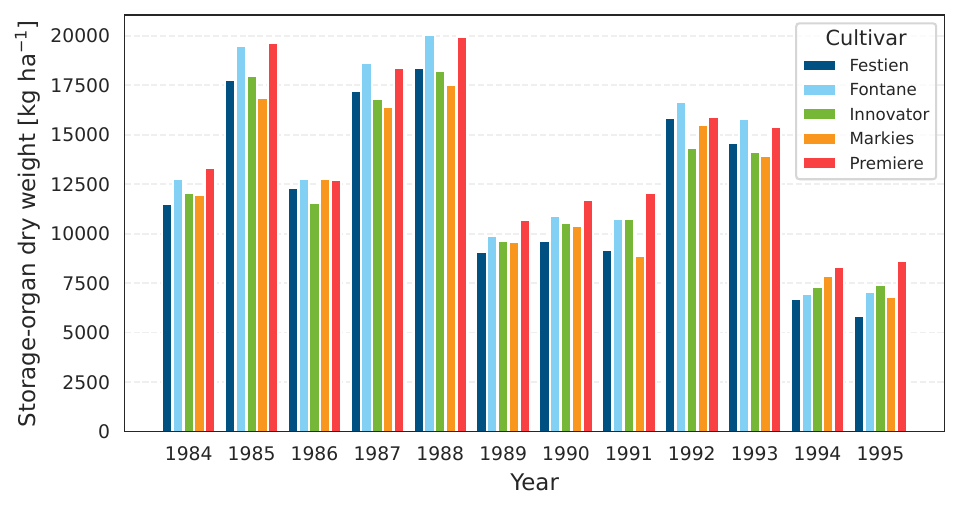}
    \caption{Dry storage organ weight (WSO) at harvest for the years and cultivars in our \textit{in silico} dataset. Within each year, cultivars are shown from left to right as Festien, Fontane, Innovator, Markies, and Premiere.}\label{fig:resulting_dataset}
    \figalttext[
        Harvest WSO for five potato cultivars across 12 simulated growing seasons.
    ]{
        Grouped bar chart showing dry storage organ weight (WSO) at harvest, in kg ha$^{-1}$, for five potato cultivars—Festien, Fontane, Innovator, Markies, and Premiere—during 12 simulated growing seasons from 1984 to 1995. WSO varies more strongly between years than between cultivars within each year. Values are highest in 1985–1988, reaching about 20,000 kg ha$^{-1}$, and lowest in 1994–1995, at roughly 6,000–8,500 kg ha$^{-1}$.
    }
\end{figure}

For each cultivar, we simulate growing seasons for the years 1984--1995 at a
fixed location (latitude N~51.0, longitude E~5.0), starting on April~20 and
ending at the earlier of the model-defined end of the phenological cycle
(\texttt{DVS}=2) or, if that is not reached in time, October~30. Daily weather
inputs are obtained from the NASA POWER database~\citep{nasa_power}. Additional
details on the simulation settings are provided in Supplementary Material.

The simulations produce daily trajectories of LAI and WSO together with
environmental variables used as ``observations'' for model identification and
validation. WOFOST soil moisture is used only to calibrate a simple proxy
water-balance model in the training environments and is not supplied directly
to SINDy in held-out environments; instead, the predicted soil-moisture proxy
is used. This WOFOST simulation dataset helps us assess the ability of SINDy to
identify parsimonious ODEs, with known reference dynamics and clean data, while
still capturing realistic environmental variation and cultivar-specific
responses.

\subsubsection{Exploratory analysis of the dataset}
\begin{figure}[H]
    \centering
    \includegraphics[width=0.70\linewidth]{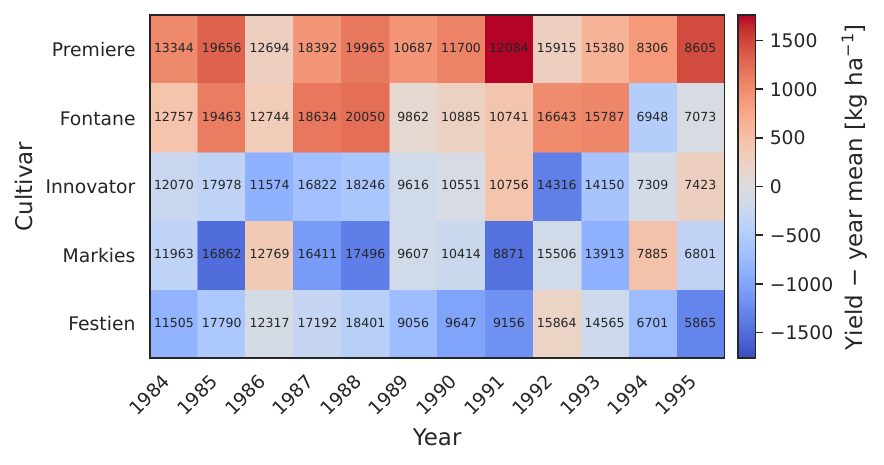}
    \caption{Year-centred yield matrix. Blue and red indicate yields below and
        above the within-year mean, respectively. Cell annotations show the corresponding (uncentered) simulated yield values.}\label{fig:yield_heatmap}
    \figalttext[
        Year-centred yield of five cultivars across 12 simulated growing seasons.
    ]{
        eat map of year-centred yield for five cultivars across 12 simulated
        growing seasons from 1984 to 1995. Cells show deviations from the mean
        yield within each year, with blue representing below-average and red
        above-average yield; printed numbers give the corresponding uncentred
        simulated yields. Relative cultivar performance varies among years,
        although Premiere is generally higher and Festien generally lower.
    }
\end{figure}

Figure~\ref{fig:resulting_dataset} shows the simulated WSO values at harvest
over the 12 years, while Figure~\ref{fig:yield_heatmap} shows the year-centred
yield matrix, where values represent deviations from the annual mean yield.
These figures indicate that variation in WSO at harvest is primarily driven by
the year (i.e., the environment), as variation across years is substantially
larger than variation across cultivars within a given year.

At the same time, the ranking of cultivars changes across years, indicating the
presence of genotype-by-environment (G$\times$E) interactions. The data show
that Premiere and Festien generally have the highest and lowest yields,
respectively, although exceptions occur in certain years.

\begin{figure}[H]
    \centering
    \includegraphics[width=0.65\linewidth]{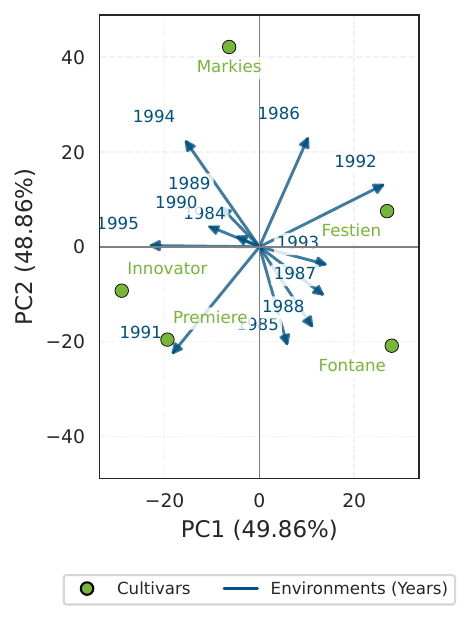}
    \caption{
        AMMI biplot of yield data showing cultivar-by-environment
        (G$\times$E) interaction. Cultivars are shown as points and years
        as vectors for the first two interaction principal components,
        using symmetric scaling ($\lambda=0.5$). PC1 and PC2 account for
        49.86\% and 48.86\% of the represented interaction variation,
        respectively.
    }\label{fig:ammi_biplot}
    \figalttext[
        AMMI biplot of cultivar-by-year yield interaction.
    ]{
        AMMI biplot showing five cultivars as points and 12 simulated
        growing seasons as vectors across the first two interaction
        principal components. Cultivars and years differ in their
        positions and directions, indicating cultivar-specific responses
        to growing seasons. Cultivars and years pointing in similar
        directions indicate positive interactions. For example, Premiere
        and 1991 are closely aligned.
    }

\end{figure}

A decomposition of the cultivar-by-year matrix confirms these observations. A
two-way decomposition of the yield data into additive cultivar and year effects
shows that the year effect accounts for $95.85\%$ of the total sum of squares,
while the cultivar effect accounts for only $2.53\%$. The remaining $1.61\%$
corresponds to cultivar-by-environment interaction. Although the contribution
of G$\times$E is small relative to the total variation, it constitutes a
substantial fraction of the variation associated with cultivar differences:
\begin{align}
    \frac{SS_{G\times E}}{SS_G + SS_{G\times E}} = 38.9\%.
\end{align}
Here, $SS_{G\times E}$ denotes the sum of squares of the
cultivar-by-environment interaction, and $SS_G$ the sum of squares of the
cultivar effect. The AMMI biplot in Figure~\ref{fig:ammi_biplot} further
illustrates the cultivar-environment interaction structure. Cultivars exhibit
different responses across years. For example, Markies shows a negative
cultivar-by-year interaction effect in 1991 and a positive one in 1986 (see
also Figure~\ref{fig:resulting_dataset}).

\subsection{Applying SINDy to the in silico dataset}\label{subsec:model_pipeline}

Successful application of the SINDy algorithm in this setting depends on an
appropriate choice of target state variables and candidate library functions.
Correlations among environmental variables (EVs) and candidate library
functions can lead to alternative ODE representations with similar predictive
performance. However, we observe that including certain types of functions can
substantially reduce the complexity of the resulting model and the number of
required GSCs. Hence, while different libraries may perform similarly in terms
of predictive error, they can differ substantially in sparsity.

\subsubsection{Selection of state and driving variables}

To model the dynamics of the two state variables, we apply SINDy to identify a
system of two ODEs of the form~\eqref{eq:SINDy-ODE-system}:
\begin{align}
    \label{eq:SL-SINDy-ODE-system}
    \begin{split}
        \frac{d S}{dt} & = \sum_{j=1}^{\ell}\xi_{j,S}\theta_{j}(t) + e_{S}(t), \qquad S(t^{S}_{0}) = S_0, \\
        \frac{d L}{dt} & = \sum_{j=1}^{\ell}\xi_{j,L}\theta_{j}(t) + e_{L}(t), \qquad L(t^{L}_{0}) = L_0,
    \end{split}
\end{align}
where $\theta_{j}(t)$ are the library functions, $\xi_{j,S}$ and $\xi_{j,L}$
are the GSCs, and $e_{S}(t)$ and $e_{L}(t)$
denote model errors. The choice of the times $t^{S}_{0}$ and $t^{L}_{0}$ is
discussed in the next section.

\begin{figure}[H]
    \centering
    \includegraphics[width=0.70\linewidth]{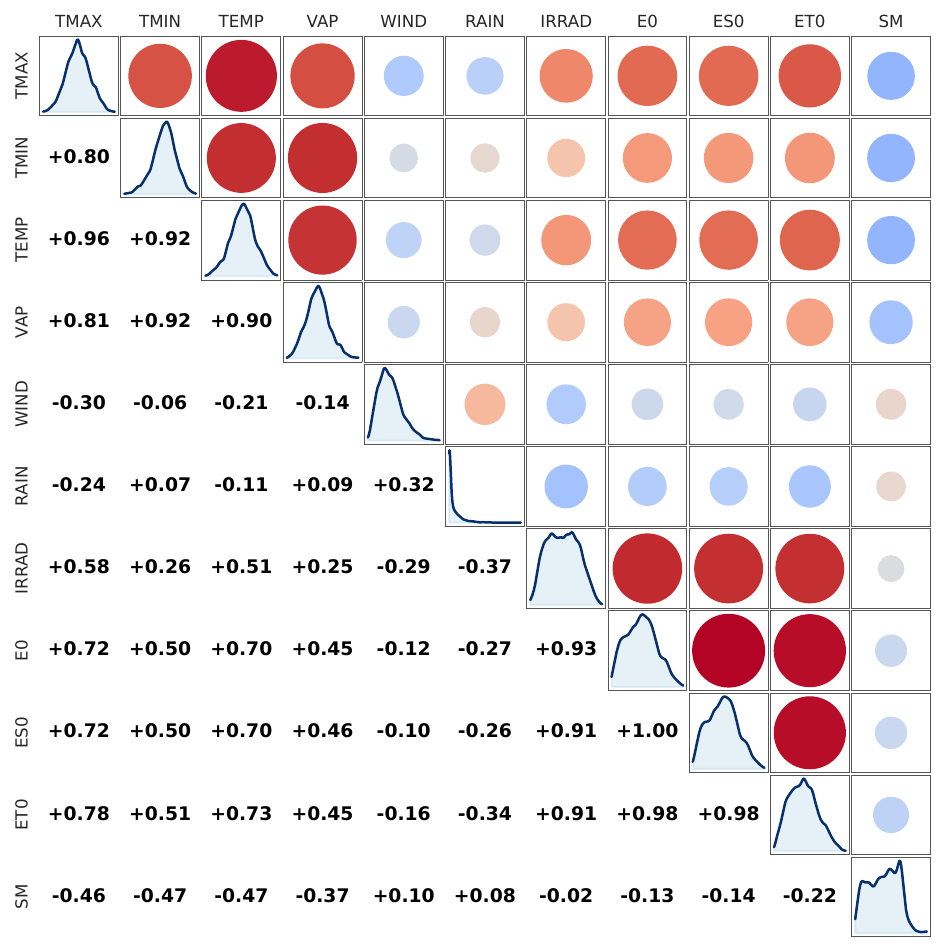}
    \caption{Correlation matrix of the 11 WOFOST environmental variables across all cultivars and years in the dataset. Off-diagonal entries show pairwise correlation coefficients, and diagonal panels show Gaussian kernel density estimates of the corresponding variable distributions. Variable definitions are provided in Table~S3.}\label{fig:correlation_matrix}
    \figalttext[Correlation matrix of the 11 WOFOST environmental variables.]{Correlation matrix of 11 environmental variables. Strong positive correlations occur among temperature-related variables and among radiation and evapotranspiration variables. Temperature and vapour pressure are generally negatively associated with soil moisture, while rainfall shows comparatively weak correlations with most variables. These relationships motivate the selection of representative environmental drivers.}
\end{figure}

To reduce redundancy among candidate functions and limit model complexity, we
select \textit{a priori} a reduced set of driving EVs to include in the library
functions $\theta_j(t)$, based on a correlation analysis of the WOFOST EVs.
Alternative data-driven feature-selection methods could also be used, but were
not explored in this work. Our feature selection is based on an analysis of
pairwise correlations between the 11 WOFOST EVs (see
Supplementary Table~S3 for an overview of the EVs and their
definitions) across all cultivars and years in our dataset (see
Figure~\ref{fig:correlation_matrix}). Several variables exhibit strong
correlations ($|r| \geq 0.9$), notably: (i) temperature-related variables
(TEMP, TMIN, TMAX, VAP), and (ii) radiation/evapotranspiration variables
(IRRAD, E0, ES0, ET0).

Most WOFOST EVs are weather-derived and therefore shared by cultivars within a
year. WOFOST soil moisture can nevertheless differ between cultivars through
crop water use. In contrast, the fitted soil-moisture proxy used below is
shared across cultivars within each year; among model inputs, only measured LAI
is cultivar-specific.

We therefore retain TEMP as the representative temperature driver, denoted $T$,
and ET0 as the representative evapotranspiration driver, denoted $E$. Wind
speed (WIND) enters WOFOST only through ET0, while rainfall (RAIN) primarily
affects soil moisture (SM). Since a proxy for soil moisture is included
explicitly (denoted as $M$, see
Section~\ref{subsubsec:soil_moisture_approximation}), we omit WIND and RAIN.
Consequently, from the original 11 EVs, we retain three primary drivers: $T$,
$E$, and $M$.

To capture dependence on phenological time, we introduce a fourth driver $I$,
defined as the cumulative sum of daily temperatures:
\begin{align}
    I(t_i)=\sum_{j=0}^{i} T(t_j).
\end{align}
In WOFOST and other process-based CGMs, phenology is often represented through
developmental-stage variables and threshold-based switches that activate or
deactivate processes. We do not include such switches explicitly. Instead, $I$
serves as a continuous proxy for phenological time. Note that $I$ differs from
conventional degree-day definitions that use $\max(T(t) - T_{\text{base}}, 0)$ (where
$T_{base}$ is a threshold temperature), since introducing such a threshold
would make $I$, and hence the candidate library, depend nonlinearly on an
additional parameter. Terms that depend on $I$ (linear or nonlinear) can
capture changes in growth dynamics across phenological stages, without
requiring explicit modelling of developmental stages.

The final set of driving variables is therefore given by
\begin{equation}
    \label{eq:DrivingEVs}
    \bm{u}(t) = [T(t), E(t), M(t), I(t)].
\end{equation}

Other environmental and management factors, such as soil properties, are kept
constant across simulations, and nitrogen limitation is not simulated. This
simplification allows us to focus on the dynamics of LAI and WSO. Incorporating
additional variables into the SINDy framework is straightforward, as they can
be included as additional columns in the driving-variable matrix $U$ (see
Equation~\eqref{eq:UmatrixSLcase}).

Given observations of $S(t)$, $L(t)$, and $\bm{u}(t)$ at time points
$t_1,\dots,t_m$, the corresponding state data matrix $Y$ is given by
\begin{align}
    \label{eq:YmatrixSLcase}
    Y =
    \begin{bmatrix}
        S(t_1) & L(t_1) \\
        S(t_2) & L(t_2) \\
        \vdots & \vdots \\
        S(t_m) & L(t_m)
    \end{bmatrix},
\end{align}
and the driving-variable matrix $U$ is given by
\begin{align}
    \label{eq:UmatrixSLcase}
    U =
    \begin{bmatrix}
        T(t_1) & E(t_1) & M(t_1) & I(t_1) \\
        T(t_2) & E(t_2) & M(t_2) & I(t_2) \\
        \vdots & \vdots & \vdots & \vdots \\
        T(t_m) & E(t_m) & M(t_m) & I(t_m)
    \end{bmatrix}.
\end{align}

\subsubsection{Start and end times and initial conditions}\label{subsubsec:start_end_times_initial_conditions}
After planting, leaf area and storage-organ biomass begin to develop at
different times, depending on cultivar and environment. We denote the onset of
active LAI and WSO development by $t_0^L$ and $t_0^S$, respectively. Before
these times, the corresponding state and growth rate are zero. Later in the
season, LAI declines due to senescence and becomes negligible at $t_1^L$. We
define $t_1^S$ as the time at which WSO growth stops, after which $\dot S(t)=0$
for all $t > t_1^S$.

Since WSO is the primary trait of interest and active WSO growth typically
starts after LAI development ($t_0^S > t_0^L$), we model both $S$ and $L$ over
the WSO growth interval $[t_0^S, t_1^S]$. However, these times are not known
\textit{a priori} for a test environment and may vary across cultivars and
years. We therefore estimate the start and end of this interval using only
information available at prediction time.

For the simulated data, reference values of $t_0^S$ and $t_1^S$ are defined
from the WOFOST trajectories as the first day for which $\dot S(t) > 5\
    \mathrm{kg\ ha^{-1}\ day^{-1}}$ and the last day before $\dot S(t)$
subsequently falls below this threshold, respectively. For a validation or test
environment $e^\ast$, let $\mathcal{D}_{\mathrm{train}}$ denote the
corresponding set of training environments. The start and end times are
predicted from the temperature integral $I(t)$ as
\begin{align}
    \hat t^{S}_{0,e^\ast}
     & =
    \min\left\{
    t :
    I_{e^\ast}(t)
    \geq
    \operatorname{median}_{e\in\mathcal{D}_{\mathrm{train}}}
    I_e(t^{S}_{0,e})
    \right\},
    \\
    \hat t^{S}_{1,e^\ast}
     & =
    \max\left\{
    t :
    I_{e^\ast}(t)
    \leq
    \operatorname{median}_{e\in\mathcal{D}_{\mathrm{train}}}
    I_e(t^{S}_{1,e})
    \right\}.
\end{align}
To ensure consistency between model identification and prediction, the same
temperature-integral-based rule is also used to define the fitting window in
each training environment. Furthermore, during cross-validation, all
thermal-time thresholds and initial-condition estimates are recomputed using
only the training environments of the corresponding fold. The predicted times
$\hat t^{S}_{0,e^\ast}$ and $\hat t^{S}_{1,e^\ast}$ define the interval over
which the SINDy model is applied in the test environment $e^\ast$.

The initial conditions $S_0$ and $L_0$ are estimated from the corresponding
training environments. Specifically, for each training environment, $S$ and $L$
are evaluated at the time point whose temperature integral is closest to
$I_{e^\ast}(\hat t^{S}_{0,e^\ast})$, after which the median values across
training environments are used as $S_0$ and $L_0$. Thermal-time thresholds and
initial-condition estimates are computed separately for each cultivar, since
phenology has a strong cultivar-specific component.

Alternatively, the initial conditions can be estimated jointly with the model
parameters, as proposed in~\citep{wei2022sparse}.

\subsubsection{Soil moisture approximation}\label{subsubsec:soil_moisture_approximation}
Directly supplying the WOFOST soil-moisture trajectory to SINDy in a test
environment would provide information from the underlying crop model that is
not generally available at prediction time. To avoid this source of data
leakage, we approximate soil moisture using a simple auxiliary water-balance
model driven by EVs. This model does not require the WOFOST soil-moisture
trajectory during the growing season and is inspired by a bucket-type
water-balance formulation with a leaky drainage term above field
capacity~\citep{arnold1998large} and an FAO-56 water-stress
response~\citep{allen1998crop}:
\begin{equation}
    \begin{split}
        \frac{dM}{dt}
        =
        {} & c_P\, P(t)
        -
        c_{\mathrm{ET}}\, K_{\mathrm{ph}}(I(t))\, K_s(M(t))\, ET_0(t) \\
        & -
        k_{\mathrm{drain}}
        \max\!\big(M(t)-M_{\mathrm{FC}},\,0\big).
    \end{split}
\end{equation}
Here, precipitation adds water to the soil reservoir, while evapotranspiration
removes water at a rate determined by reference evapotranspiration, a
thermal-time-based approximation of crop development $K_{\mathrm{ph}}$, and a
soil-moisture stress function $K_s$. Drainage removes water when soil moisture
exceeds the field-capacity threshold $M_{\mathrm{FC}}$. The complete
water-balance formulation, including the functional forms of
$K_{\mathrm{ph}}$ and $K_s$, is provided in the supplementary material.

The water-balance model contains seven unknown parameters. These parameters are
estimated using only the cultivar-averaged WOFOST soil-moisture trajectories
from the corresponding training environments. For a validation or test
environment, the model requires only a single soil-moisture value at planting,
$M(t_p)$, which is treated as an initial condition available at prediction
time. In the present simulation experiment, this input is fixed to the known
field-capacity value $M(t_p)=0.30$~cm$^3$~cm$^{-3}$ used to initialise each
WOFOST simulation. It is therefore available at planting and is not extracted
from the held-out WOFOST soil-moisture trajectory. No subsequent WOFOST
soil-moisture values from that environment are used.

To ensure consistency between model identification and prediction, the
resulting approximation $\hat M(t)$ is used as the soil-moisture input both
when fitting SINDy in the training environments and when evaluating the model
in validation or test environments. Within cross-validation, all water-balance
parameters are re-estimated using only the training environments of the
corresponding fold. This prevents the SINDy equations from being fitted with
WOFOST soil-moisture information that would be unavailable when predicting a
held-out year.

\subsubsection{Library construction}\label{subsubsec:library_construction}
Having defined the data and driving-variable matrices $Y$ and $U$ in
Equation~\ref{eq:YmatrixSLcase} and Equation~\ref{eq:UmatrixSLcase}, we
construct the matrix $\Theta$ (see Supplementary Equation~S7) of
sampled library functions for a polynomial library of degree $p$, as
illustrated in Supplementary Equation~S5. In this
work, we consider polynomial libraries with varying degrees $p$ and
compositions, as these are commonly used in the SINDy
literature~\citep{brunton2016discovering}. The choice of $p$ represents a
trade-off between model complexity and flexibility.

\begin{figure}[H]
    \centering
    \includegraphics[width=\linewidth]{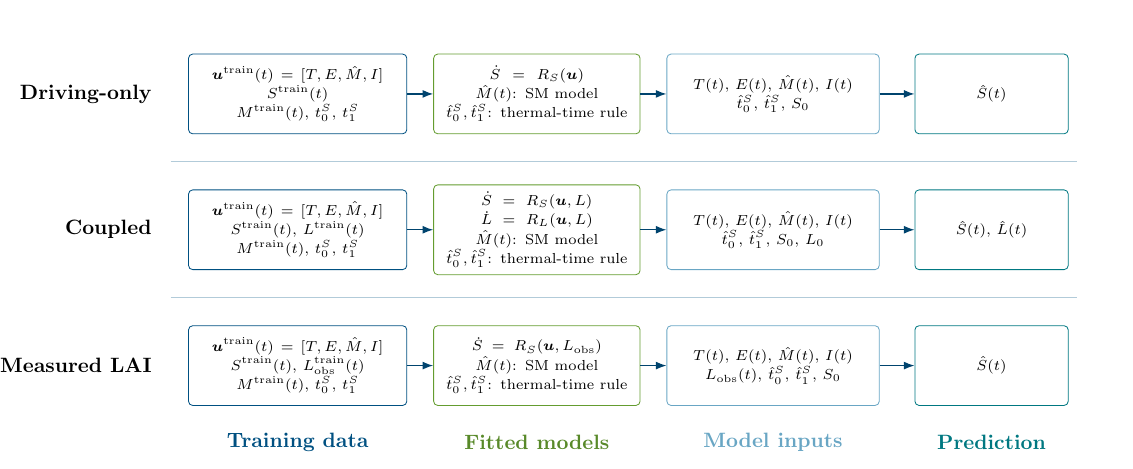}
    \caption{Schematic overview of the three model formulations and the information flow from training to prediction. Each row proceeds from training data, through fitted auxiliary models and model inputs, to the resulting prediction. Training data are used to identify the SINDy model and to fit the soil-moisture and thermal-time models. At prediction time, the models receive the environmental drivers $[T(t),E(t),\hat{M}(t),I(t)]$ and the required initial conditions and are integrated over the predicted active-growth interval $[\hat{t}^{S}_{0},\hat{t}^{S}_{1}]$. The Driving-only formulation predicts WSO without LAI information, the Coupled formulation jointly predicts WSO and LAI, and the Measured-LAI formulation uses externally observed LAI as an additional input.}\label{fig:library_compositions}
    \figalttext[Information flow for the Driving-only, Coupled, and Measured-LAI model formulations.]{Schematic of three parallel modelling workflows arranged from left to right as training data, fitted models, model inputs, and prediction. The Driving-only workflow predicts WSO from environmental drivers without LAI information. The Coupled workflow jointly predicts WSO and LAI. The Measured-LAI workflow predicts WSO using environmental drivers and externally observed LAI. In all workflows, training-derived soil-moisture and thermal-time models provide soil moisture and the predicted active-growth interval used for prediction.}
\end{figure}

We consider the following three library compositions, which we found to perform
well for modelling WSO dynamics under the temperature- and water-limited
conditions of our simulations. The approaches are also summarised schematically
in Figure~\ref{fig:library_compositions}.

\begin{enumerate}
    \item \textbf{\em Driving-only case}: The library contains only products of driving variables, including self-products, up to degree $p$. This corresponds to the assumption that WSO dynamics can be approximated solely
          in terms of the driving variables, without feedback from the state variables
          $S$ and $L$. The columns of $\Theta$ are constructed from the columns of $U$ by taking entrywise products involving at most $p$ factors. For example, for $p=2$, the library may include $T$, $T^2$, $TM$, etc., but not $T^2M$, as the latter involves three factors. The constant term (unity column) is included only once.

    \item \textbf{\em Coupled case}: Since leaf area affects the growth of
          storage organs, but not vice versa to the same extent, coupling between the
          equations in Equation~\ref{eq:SINDy-ODE-system} is assumed to be one-sided.
          Specifically, both rate functions may depend on $L$ in addition to the driving
          variables, but not on $S$. During model fitting, the observed time samples of
          $L$ are used to construct the library matrix $\Theta$. During prediction,
          however, $L(t)$ is not supplied externally but is obtained by integrating its
          own ODE together with the ODE for $S(t)$. Products involving $L$ are allowed,
          provided that only the first power of $L$ is used. For example, with $p=2$,
          terms such as $LT^2$, $LTE$, and $T^2$ are included, whereas $L^2T$ is
          excluded.

    \item \textbf{\em Measured-LAI case}: If $L(t)$ is available from measurements
          or from a proxy such as remote-sensing data, the system in
          Equation~\ref{eq:SINDy-ODE-system} can be reduced to a single equation for $S$, with
          the rate function optionally depending on $L$. The case where $S$ does not
          depend on $L$ is already covered by the driving-only case. The matrix
          $\Theta$ is constructed using the same candidate products involving $L$ as in
          the coupled case, but $L(t)$ is treated as an externally measured input both
          during fitting and during evaluation. Such a model is suitable when measurements of $L(t)$ are available over the
          period of interest, but it is not a stand-alone forecasting model because it
          does not predict $L(t)$ itself.
\end{enumerate}

Each approach has its advantages and limitations. The driving-only case returns
an uncoupled ODE for $S$, allowing model construction even when no data for $L$
are available. However, this approach does not explicitly capture the
dependence of $S$ on $L$. As a result, accurate models may require higher-order
terms in the driving variables, which reduces interpretability.

The coupled case explicitly represents the dependence of $S$ on $L$, but
requires observations of both variables during model identification. Once the
GSCs are determined, the model can be applied in a new environment without
external measurements of $L$, because $L$ is predicted jointly with $S$.
Including $L$ may also allow the dependence of $S$ on the environmental drivers
to be represented with lower-order terms.

The measured LAI case is motivated by the use of high-throughput phenotyping
data, such as drone-based canopy measurements. In this setting, the LAI state
variable $L$ can be replaced by a proxy derived from remote sensing, allowing
integration of observational data into the model for $S$.

\subsubsection{Numerical differentiation and integration}
To apply SINDy, the time derivatives $\dot S(t)$ and $\dot L(t)$ must be
estimated from the observed state trajectories. We use Savitzky–Golay
differentiation~\citep{savitzky1964smoothing}, which is well suited for smooth
time series. For the simulated daily data, we use a window size of 5
observations (i.e., 5 days) and a polynomial order of 2. These parameters
should be adjusted depending on the noise level in the data, with larger window
sizes typically required for noisier observations. Sensitivity to these
parameters is examined in the Supplementary Material. In real
experiments, where measurements are often sparser (e.g., 7–15 time points per
season), derivative estimation can be substantially more challenging and may
require dedicated methods. Weak~\citep{messenger2021weak} or
integral~\citep{wei2022sparse} formulations of SINDy can be used to deal with
sparse and noisy observations, but are not evaluated in this study.

Predicted SINDy trajectories are obtained by using a fixed-step classical
fourth-order Runge--Kutta (RK4) scheme~\citep{leveque2007finite}, with two
integration steps per day. Environmental drivers are linearly interpolated
between daily observations. If a candidate model diverges during integration
(which can occur only for the coupled case), the corresponding hyperparameter
combination is discarded.

\subsubsection{SSR optimizer}
The next step in the SINDy pipeline is to determine a sparse coefficient matrix
$\Xi$ in Supplementary Equation~S7, i.e., to fit $\dot Y \approx
    \Theta(Y,U)\,\Xi$ using as few nonzero coefficients as possible. We use
Stepwise Sparse Regression (SSR)~\citep{boninsegna2018sparse}, which
iteratively eliminates library terms while refitting the remaining
coefficients. The target model complexity is specified by $\bm{\chi}$, i.e.,
the desired number of nonzero coefficients per ODE. This hyperparameter will be
selected using cross-validation.

\begin{algorithm}[H]
    \caption{Shared-support stepwise sparse regression (SSR)~\citep{boninsegna2018sparse} across genotypes}\label{alg:ssr_shared_support}
    \begin{algorithmic}[1]
        \Require For each genotype $g\in\{1,\dots,G\}$: library matrix $\Theta^{(g)}\in\mathbb{R}^{m_g\times \ell}$ and target trait vector $\dot y_i^{(g)}\in\mathbb{R}^{m_g}$ (for a fixed trait index $i$)
        \Require Target support size $k$, ridge parameter $\alpha$, final ridge $\alpha_{\mathrm{final}}$ (typically $0$)
        \State $\mathcal S \gets \{1,\dots,\ell\}$ \Comment{active library indices}

        \Statex \textit{(Column-wise normalisation per genotype)}
        \For{$g=1,\dots,G$}
        \For{$j=1,\dots,\ell$}
        \State $c^{(g)}_j \gets \max\!\left(\|\Theta^{(g)}_{:,j}\|_2,\,1\right)$
        \State $\widetilde{\Theta}^{(g)}_{:,j} \gets \Theta^{(g)}_{:,j}/c^{(g)}_j$
        \EndFor
        \EndFor

        \While{$|\mathcal S|>k$}
        \Statex \textit{(Fit ridge on the current shared support)}
        \For{$g=1,\dots,G$}
        \State $\widehat{\xi}^{(g)}_{\mathcal S}\gets
            \arg\min_{\beta\in\mathbb{R}^{|\mathcal S|}}
            \big\|\widetilde{\Theta}^{(g)}_{:,\mathcal S}\beta-\dot y_i^{(g)}\big\|_2^2+\alpha\|\beta\|_2^2$
        \State $\widehat{\xi}^{(g)}_{\mathcal{S}^c} \gets 0$
        \EndFor

        \Statex \textit{(Compute a cross-genotype importance score per term)}
        \ForAll{$j\in \mathcal S$}
        \State $s_j \gets \left(\sum_{g=1}^{G}\big(\widehat{\xi}^{(g)}_{j}\big)^2\right)^{1/2}$
        \EndFor
        \State $j^\star \gets \arg\min_{j\in \mathcal S} s_j$
        \State $\mathcal S \gets \mathcal S\setminus\{j^\star\}$ \Comment{eliminate term with smallest cross-genotype importance}
        \State $\mathcal S^c \gets \{1,\dots,\ell\}\setminus \mathcal S$ \Comment{inactive library indices}
        \EndWhile

        \Statex \textit{(Final refit on the selected support and rescale back)}
        \For{$g=1,\dots,G$}
        \State $\bar{\xi}^{(g)}_{\mathcal S}\gets
            \arg\min_{\beta\in\mathbb{R}^{|\mathcal S|}}
            \big\|\widetilde{\Theta}^{(g)}_{:,\mathcal S}\beta-\dot y_i^{(g)}\big\|_2^2+\alpha_{\mathrm{final}}\|\beta\|_2^2$
        \ForAll{$j\in\mathcal S$}
        \State $\xi^{(g)}_j \gets \bar{\xi}^{(g)}_j/c^{(g)}_j$ \Comment{undo normalisation}
        \EndFor
        \State $\xi^{(g)}_{\mathcal{S}^c} \gets 0$
        \EndFor
        \State \textbf{Return:} Shared support $\mathcal S$ and genotype-specific coefficient vectors $\{\xi^{(g)}\}_{g=1}^G$ for trait $y_i$.
    \end{algorithmic}
\end{algorithm}

During intermediate refits, we use ridge regression with a small regularization
parameter $\alpha$ to improve numerical stability and mitigate
multicollinearity among library functions. This is particularly important for
higher-degree polynomial libraries, where many terms are correlated. Once the
support size reaches $\bm{\chi}$, a final refit is performed on the selected
terms, without regularization (OLS, i.e., $\alpha_{\mathrm{final}} = 0$), to
reduce shrinkage bias.

At each SSR iteration, we (i) estimate the coefficients on the current active
set of library terms, (ii) remove the weakest term, and (iii) repeat until the
desired model complexity is reached. To prevent effects with large magnitudes
from dominating the elimination step, we normalise the columns of the library
matrix $\Theta$ using their $\ell_2$ norms, and rescale the final coefficients
afterward.

As an adaptation of SSR to the multi-genotype setting, we enforce a shared
model structure across genotypes, inspired by group-sparsity approaches in
multi-task learning, for example~\citep{argyriou2006multi,obozinski2010joint}.
Specifically, we assume that all genotypes share the same active set of library
terms, while allowing the associated coefficients to be genotype-specific. Note
that this is analogous to the assumption underlying CGM-WGP approaches, where
the same model structure is used for all genotypes, but the GSPs (or in our
case, GSCs) are genotype-specific.

At each SSR iteration, we compute, for every candidate library term, the
$\ell_2$ norm of its coefficient vector across genotypes. The library term with
the smallest cross-genotype norm is then removed from the active set for all
genotypes. This leads to a shared-support model: the same terms are selected
for all genotypes, but the coefficient values can differ. In other words, the
same terms define the dynamical structure for every genotype, whereas the
coefficient values are genotype-specific. This shared-support formulation
reflects the modelling assumption that the cultivars can be represented by the
same functional structure, while differing in their responses to environmental
and physiological drivers. The complete procedure is described in
Algorithm~\ref{alg:ssr_shared_support}.

\subsubsection{Hyperparameter tuning and model selection}
To select the best model complexity for the given data, several hyperparameters
must be specified in the SINDy pipeline. In this work, some of these are fixed
\textit{a priori}, while others are selected via cross-validation. The
hyperparameters include:
\begin{itemize}
    \item The polynomial degree $p$ of the basis functions in the library $\Theta(Y, U)$.
    \item The ridge regularization parameter $\alpha$ used in the SSR algorithm for
          numerical stability.
    \item The desired model complexity $\bm{\chi}$, i.e., the number of non-zero
          coefficients in the ODEs for $L$ and $S$, with $\bm{\chi} = (\chi_L,
              \chi_S)^T$.
\end{itemize}
To limit the size of the hyperparameter search space, the ridge parameter is
fixed to $\alpha = 10^{-6}$ (after column normalisation). This value is only used
to stabilize the SSR elimination procedure. After the active set is determined,
the final set of coefficients is refitted without regularization (OLS). Hence,
only the model complexity $\bm{\chi}$ is selected via cross-validation. To
assess generalisation performance across years, we employ a nested
leave-one-year-out cross-validation (LOYO-CV) scheme.

Let $N$ denote the number of years in the dataset. In the outer loop, each year
$r \in \{1,\dots,N\}$ is held out in turn as the test set, while the remaining
$N-1$ years form the training set. The test year's crop state data are not used
during hyperparameter selection, model fitting, or estimation of the timing of
the active-growth window, initial-condition, and soil-moisture submodels.

Within each outer fold, an inner LOYO-CV procedure is used to select the model
complexity $\bm{\chi}^\star$. For each candidate $\bm{\chi}$, the mean
validation RMSE and its standard error are computed across the $N-1$ inner
folds. We then apply the one-standard-error (1-SE)
rule~\citep{hastie2009elements}: among all candidate models whose mean
validation RMSE is within one standard error of the minimum, we select the
model with the lowest number of non-zero coefficients. This favours more
parsimonious models when several complexities achieve similar predictive
performance.

Within each inner fold, the temperature-integral thresholds used to determine
the active growth window, the initial conditions $S_0$ and $L_0$, and the
parameters of the soil-moisture approximation are re-estimated using only the
corresponding inner-training environments before evaluation on the held-out
validation year.

Using $\bm{\chi}^\star$, a final model is trained on all $N-1$ training years
and evaluated on the held-out test year $r$. Repeating this procedure for all
outer folds yields $N$ test scores $\{T_r\}_{r=1}^N$, which are then averaged
to obtain the overall generalisation performance. Due to strong intra-annual
correlations in weather variables, cross-validation is performed between years
rather than within years.

The complete procedure is summarised in~Algorithm~\ref{alg:nested_loocv}. The
hyperparameter grid is given by $\mathcal{K} = \{\bm{\chi}^{(1)}, \dots,
    \bm{\chi}^{(K)}\}$. To keep the grid tractable, we select 10 values of $\chi_L$
and $\chi_S$ approximately evenly spaced between 1 and the maximum number of
available library functions, which depends on the polynomial degree $p$ and the
library structure. This results in $K=100$ candidate settings for the coupled
library and 10 settings for the driving-only and measured LAI libraries. Model
performance is quantified using the root mean squared error (RMSE) of the
predicted storage organ dry weight $S$. The 1-SE rule provides an explicit
accuracy--complexity trade-off by favouring lower-complexity models when
several candidate complexities achieve similar predictive performance. The
selected support should therefore be regarded as a compact predictive
representation, rather than as a uniquely or maximally parsimonious
representation.

This procedure is repeated for each polynomial degree $p \in \{1,\dots,4\}$ and
for each library structure (see Section~\ref{subsubsec:library_construction}).
Higher degrees are not considered due to the combinatorial growth of the
library size.

\begin{algorithm}[H]
    \caption{Nested leave-one-out cross-validation for hyperparameter tuning and model selection. $C(\bm{\chi})$ denotes the complexity of the model with hyperparameters $\bm{\chi}$, $\mathrm{SE}(\bm{\chi})$ denotes the standard error of the validation RMSE for that model.}\label{alg:nested_loocv}
    \begin{algorithmic}[1]
        \Require Year-indexed dataset $\mathcal{D}=\{\mathcal{D}_r\}_{r=1}^{N_{\text{years}}}$, where each $\mathcal{D}_r$ contains the time-series data required by the SINDy--SSR pipeline
        \Require Fixed ridge parameter $\alpha$, complexity grid $\mathcal{K}$ (candidate values of model complexity $\bm{\chi}$)

        \For{$r=1$ \textbf{to} $N_{\text{years}}$}
        \State $\mathcal{D}_{\text{test}} \gets \mathcal{D}_r$
        \State $\mathcal{D}_{\text{train}} \gets \mathcal{D}\setminus\{\mathcal{D}_{\text{test}}\}$

        \For{each complexity setting $\bm{\chi} \in \mathcal{K}$}
        \State Initialise validation loss $J_{\bm{\chi}} \gets 0$
        \For{each year index $v \in \{1,\dots,N_{\text{years}}\}\setminus\{r\}$}
        \State $\mathcal{D}_{\text{val}} \gets \mathcal{D}_v$
        \State $\mathcal{D}_{\text{tr}} \gets \mathcal{D}_{\text{train}}\setminus\{\mathcal{D}_{\text{val}}\}$
        \State $\mathcal{M}_{\bm{\chi},v} \gets \text{SSR}(\mathcal{D}_{\text{tr}};\,\bm{\chi},\alpha)$
        \State $j_{\bm{\chi},v} \gets \text{Loss}(\mathcal{M}_{\bm{\chi},v},\,\mathcal{D}_{\text{val}})$
        \State $J_{\bm{\chi}} \gets J_{\bm{\chi}} + j_{\bm{\chi},v}$
        \EndFor
        \State $J_{\bm{\chi}} \gets J_{\bm{\chi}}/(N_{\text{years}}-1)$
        \Comment{average over $N_{\text{years}}-1$ inner folds}
        \EndFor

        \State $\bm{\chi}_{\min} \gets \arg\min_{\bm{\chi}\in\mathcal{K}} J_{\bm{\chi}}$
        \State $\bm{\chi}^\star \gets
            \arg\min_{\bm{\chi}\in\mathcal{K}}
            \left\{
            C(\bm{\chi}) :
            J_{\bm{\chi}} \le
            J_{\bm{\chi}_{\min}} + \mathrm{SE}_{\bm{\chi}_{\min}}
            \right\}$

        \State $\mathcal{M}_r \gets \text{SSR}(\mathcal{D}_{\text{train}};\,\bm{\chi}^\star,\alpha)$
        \State $T_r \gets \text{Loss}(\mathcal{M}_r,\,\mathcal{D}_{\text{test}})$
        \EndFor

        \State \textbf{Return:} Test losses $\{T_r\}_{r=1}^{N_{\text{years}}}$ and corresponding selected models $\{\mathcal{M}_r\}_{r=1}^{N_{\text{years}}}$.
    \end{algorithmic}
\end{algorithm}

In addition to providing an out-of-sample estimate of the model's performance
on held-out years, this procedure also enables the assessment of the
variability of the model structure and performance across different years. This
provides insight into the robustness of the identified model structure and
motivates the subsequent analyses of support, coefficient, and functional
stability across different environmental conditions.

\subsubsection{Baseline models}

To benchmark the predictive performance of the SINDy models against alternative
approaches, we consider several baseline models of varying complexity. Baseline
models are fitted separately for each cultivar and evaluated using the same
outer LOYO-CV splits and prediction-time information as the SINDy models. For
every outer and inner split, the active-growth timing and soil-moisture proxy
models are estimated from the corresponding training environments only, as in
the SINDy procedure. Here, we give a brief overview of the baseline models;
further details are provided in Supplementary Material.

These baselines address complementary questions: whether our dynamic modelling
approach improves on static yield prediction, how much environment-driven
models improve over simple trajectory shapes, how much accuracy we lose or gain
by promoting sparsity, and how SINDy compares with a flexible nonlinear dynamic
model. All models are compared using the same held-out LOYO-CV splits.

As simple trajectory baselines, we construct cultivar-specific mean growth
trajectories from the training years alone, aligned by temperature integral
$I$. We also fit cultivar-specific Gompertz growth
curves~\citep{yin2003flexible}, using temperature integral $I$ as the
independent variable. For both models, predicted yield is defined as the
maximum predicted WSO over the predicted growth window. As a corresponding
simple static baseline, we use the cultivar-specific mean yield over the
training years.

For direct final-yield prediction, we additionally fit static Ridge-regression
models~\citep{hoerl1970ridge} based on summaries of conditions within the
predicted growth window. These include window timing and duration, and
cumulative and distributional summaries of temperature, reference
evapotranspiration, precipitation, temperature integral, and proxy soil
moisture. A second Ridge model additionally includes summaries of the observed
LAI trajectory.

To isolate the effect of sparse model selection, we fit dense polynomial ODE
models using the same candidate-function libraries as SINDy, but estimate all
library coefficients using Ridge regression. These models are evaluated for
polynomial degrees $p=1,\ldots,4$ and for the Driving-only, Measured-LAI, and
Coupled formulations.

Finally, we fit a neural ordinary differential equation (Neural ODE)
model~\citep{chen2018neural} as a flexible nonlinear dynamic baseline. For each
cultivar, we represent the growth-rate function by a single fully connected
network with two hidden layers. We compare the same three formulations as for
SINDy: Driving-only, Measured-LAI, and Coupled. For each formulation, the
network takes as input the driving variables and, where applicable, the
formulation-specific LAI input (either measured or predicted). For the
Driving-only and Measured-LAI formulations, the network predicts the scalar WSO
growth rate $\dot S$; for the Coupled formulation, the same network jointly
outputs both $\dot S$ and $\dot L$, rather than fitting the two rates with
separate models. Parameters are optimised by minimising the mean squared error
between the observed and network-simulated trajectories over the full training
window, obtained by numerically integrating $f_\theta$ with a fixed step RK4
scheme (two substeps per day) used for the SINDy and Ridge ODE rollouts.

\subsubsection{Model validation and accuracy}
\label{subsec:model_validation_and_accuracy}

To assess predictive performance, we evaluate storage-organ dry weight $S$
dynamically over the growing season and at harvest. Seasonal trajectory metrics
are computed over the predicted active-growth interval,
$[\hat{t}^{S}_{0,e^\ast},\hat{t}^{S}_{1,e^\ast}]$, as defined in
Section~\ref{subsubsec:start_end_times_initial_conditions}.

Because the predicted $S$ trajectories are not constrained to be monotonic and
may show declines or negative values, we consider two definitions of predicted
yield: (i) the maximum predicted $S$ over the active-growth interval, and (ii)
the predicted $S$ at the end of the active-growth period,
$\hat{t}^{S}_{1,e^\ast}$. The first definition is used for the main results,
while the second is compared in Supplementary Material.

For the coupled formulation, errors in the predicted LAI $L$ can propagate to
the WSO prediction through the coupled ODE system. We therefore also report the
predictive accuracy of $L$ for the coupled models.

Model performance is quantified using the root mean squared error (RMSE), mean
absolute error (MAE), and coefficient of determination ($R^2$). Definitions of
these metrics are provided in Supplementary Material.

\section{Numerical Experiments}\label{sec:results}
In this section, we present a series of numerical experiments evaluating the
SINDy framework on the \textit{in silico} dataset. We first assess the soil
moisture and active growth period approximations that are required to apply the
identified models in held-out environments. We then examine how library
composition and polynomial degree affect predictive performance and model
complexity, and compare the identified models with alternative dynamic
baselines. Next, we investigate the stability of the identified model
structures and learned rate functions, and analyse their functional
interpretation. We subsequently evaluate whether the models capture variation
in end-of-season yield across environments and cultivars, including
year-centred and G$\times$E-related components of yield variation. Finally, we
perform an illustrative sensitivity analysis to observation noise and sampling
density.

\subsection{Evaluation of prediction-time approximations}
Before evaluating the SINDy models, we assess the approximations required to
apply them in held-out environments without access to the corresponding WOFOST
crop trajectories. Specifically, we evaluate the predicted start and end of the
active growth period and the weather-driven water-balance approximation of soil
moisture. These results are presented Figure~\ref{fig:soil_moisture_proxy} and
Table~\ref{tab:oracle_ablation}, with additional details in Supplementary
Material.

\begin{figure}[H]
    \centering
    \includegraphics[width=0.75\linewidth]{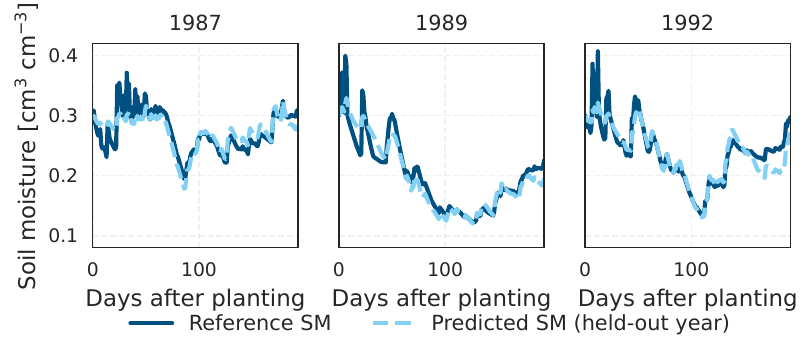}
    \caption{Reference and predicted soil moisture for three held-out years (1987, 1989, and 1992). Reference soil moisture is obtained from WOFOST simulations and averaged over cultivars. Predicted soil moisture is obtained from the weather-driven water-balance model calibrated using the remaining 11 years.}\label{fig:soil_moisture_proxy}
    \figalttext[Reference and predicted root-zone soil moisture in three held-out years.]{Three line plots for 1987, 1989, and 1992 compare WOFOST reference root-zone soil moisture with predictions from the weather-driven water-balance model. Predicted soil moisture generally follows the seasonal pattern of the reference values, with some limited differences in timing and magnitude.}
\end{figure}

The temperature-integral procedure predicts the start of the active growth
period accurately, with an MAE of $1.3$ days, whereas prediction of the end of
the active period is less precise, with an MAE of $7.1$ days (averaged across
all cultivars and held-out years). Prediction bias is small in both cases:
$+0.4$ days for the start and $-2.0$ days for the end. As shown by the
oracle-information ablation in Table~\ref{tab:oracle_ablation}, however,
uncertainty in the predicted growth window has little effect on WSO prediction.

The weather-driven water-balance model reproduces the main seasonal variation
in WOFOST root-zone soil moisture (Figure~\ref{fig:soil_moisture_proxy}).
Across the 12 held-out years, the predicted soil-moisture trajectories achieve
an RMSE of $0.0187$ cm$^3$ cm$^{-3}$ and $R^2=0.91$.

\begin{table}[th]
    \centering
    \caption{Oracle-information ablation for a measured-LAI SINDy model with $p=3$. ``True'' soil moisture and growth windows use
        WOFOST-derived values, whereas ``predicted'' values use the
        approximations described in
        Sections~\ref{subsubsec:soil_moisture_approximation}
        and~\ref{subsubsec:start_end_times_initial_conditions}.
        Entries report test WSO RMSE (mean $\pm$ SD across the 60 cultivar--year combinations in the held-out years), kg ha$^{-1}$.}\label{tab:oracle_ablation}
    \begin{tabular}{lcc}
\toprule
 & \textbf{Predicted window} & \textbf{True window} \\
\midrule
\textbf{Predicted SM} & 527.8 $\pm$ 337.4 & 502.0 $\pm$ 395.6 \\
\textbf{True SM} & 480.1 $\pm$ 216.0 & 471.6 $\pm$ 230.0 \\
\bottomrule
\end{tabular}

\end{table}
To quantify how these prediction-time approximations affect SINDy performance,
we perform an oracle-information ablation using a measured-LAI model with
polynomial degree $p=3$. We independently replace the predicted growth window
and predicted soil-moisture trajectory by their WOFOST-derived counterparts
(Table~\ref{tab:oracle_ablation}). Replacing the predicted growth window has a
modest and condition-dependent effect on WSO prediction. With predicted soil
moisture, mean RMSE decreases from $527.8$ to $502.0$ kg ha$^{-1}$ when the
WOFOST-derived growth window is used. With WOFOST soil moisture, the
corresponding RMSE changes from $480.1$ to $471.6$ kg ha$^{-1}$. In contrast,
replacing the predicted soil-moisture trajectory by WOFOST soil moisture
improves prediction accuracy, reducing mean RMSE from $527.8$ to $480.1$ kg
ha$^{-1}$ when the predicted growth window is retained.

These results show that the prediction-time approximations do not account for
the majority of the SINDy prediction error. In particular, replacing the
predicted active-growth window by the WOFOST-derived window has little effect
on WSO prediction, whereas access to the WOFOST soil-moisture trajectory
provides a moderate improvement. We therefore use the predicted growth window
and weather-driven soil-moisture trajectory in all subsequent held-out-year
evaluations. More detailed results of the oracle-information ablation are
provided in the Supplementary Material.

\subsection{Influence of library composition and polynomial degree}

\begin{figure}[H]
    \begin{minipage}[t]{0.45\textwidth}
        \centering
        \includegraphics[width=\linewidth]{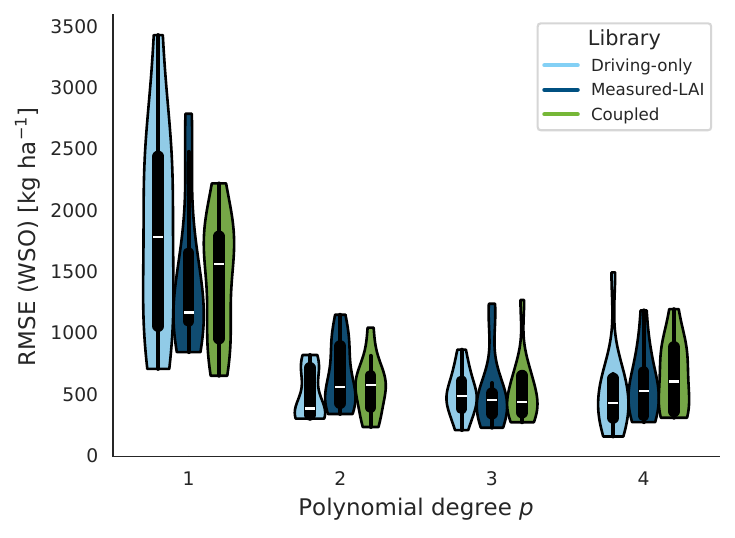}
        \textbf{(a)} Violin plots of the test RMSE in $S$ for different degrees and library compositions, across all outer folds for all five cultivars.
    \end{minipage}
    \hfill
    \begin{minipage}[t]{0.45\textwidth}
        \centering
        \includegraphics[width=\linewidth]{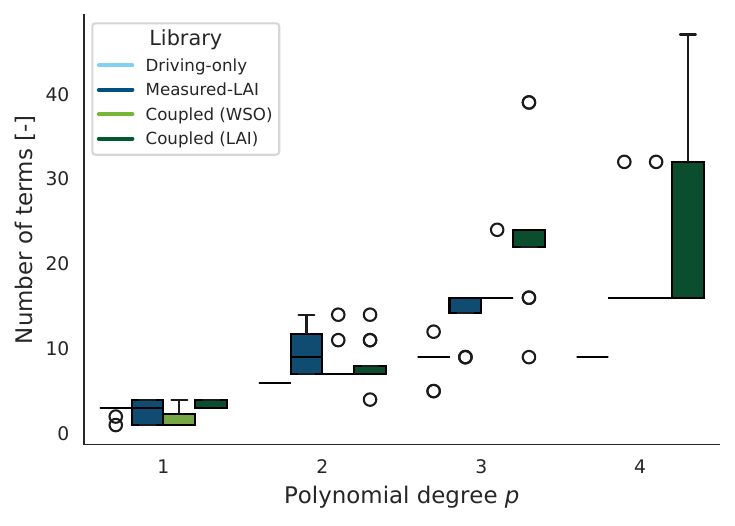}
        \textbf{(b)} Selected model complexity for different library compositions and polynomial degrees across the outer folds.
    \end{minipage}
    \vspace{0.7em}
    \caption{Predictive performance and selected model complexity across polynomial degrees and library compositions. \textbf{(a)} Test root mean squared error (RMSE) of storage-organ dry weight $S$ (WSO) across the outer folds and five cultivars. Within each polynomial degree, the Driving-only, Measured-LAI, and Coupled formulations are shown from left to right. \textbf{(b)} Selected model complexity, expressed as the number of selected terms across the outer folds. Within each degree, Driving-only, Measured-LAI, Coupled ($S$), and Coupled ($L$) are shown from left to right. The Coupled formulation fits separate ODEs for $S$ and $L$, so the complexity of both equations is shown separately. In these simulations, $S$ ranges from 0 early in the growing season to approximately 20,000~kg~ha$^{-1}$ near harvest.}\label{fig:violin_and_boxplots}
    \figalttext[Prediction error and selected model complexity across polynomial degrees.]{Two-panel figure comparing predictive error and model complexity for polynomial degrees one to four. Panel (a) shows test RMSE distributions for the Driving-only, Measured-LAI, and Coupled formulations. Degree one produces substantially larger and more variable errors, while degrees two to four generally have lower errors. Panel (b) shows the number of selected terms for Driving-only, Measured-LAI, and the separate WSO and LAI equations of the Coupled formulation. Model complexity generally increases with polynomial degree and is largest and most variable at degree four.}
\end{figure}

We evaluate predictive performance across library compositions and polynomial
degrees using the outer-loop test RMSE from nested LOYO-CV (leave-one-year-out
cross-validation) (Table~\ref{tab:summary_hyperparameter_results}).
Figure~\ref{fig:violin_and_boxplots} provides a visual summary of these
results, while cultivar-specific results are reported in Supplementary
Material.

\begin{table}[h]
    \caption{Test RMSE (mean $\pm$ sample SD) of storage organ dry weight $S$ [kg ha$^{-1}$]
        across 60 held-out cultivar--year trajectories, for different polynomial degrees and library
        compositions. Cultivar-specific results are shown in Supplementary
        Material. Bold entries indicate
        the lowest RMSE for each degree across library compositions.}
    \label{tab:summary_hyperparameter_results}
    \centering
    \begin{tabular}{c c c c}
\toprule
\textbf{Degree} & \textbf{Driving-only} & \textbf{Measured-LAI} & \textbf{Coupled} \\
\midrule
1 & 1825.5 $\pm$ 917.9 & 1434.5 $\pm$ 670.6 & \textbf{1426.3 $\pm$ 580.1} \\
2 & \textbf{493.8 $\pm$ 245.8} & 644.5 $\pm$ 331.1 & 555.2 $\pm$ 278.5 \\
3 & \textbf{499.4 $\pm$ 230.4} & 527.8 $\pm$ 337.4 & 523.1 $\pm$ 333.9 \\
4 & \textbf{513.2 $\pm$ 375.9} & 575.7 $\pm$ 315.1 & 639.9 $\pm$ 383.0 \\
\bottomrule
\end{tabular}

\end{table}

\begin{table}[b]
    \caption{Median selected model complexity (number of terms) across outer folds,
        for different polynomial degrees and library compositions. $k/K$ denotes
        the median number of selected terms $k$ out of the total number of candidate
        terms $K$.}
    \label{tab:model_complexity_summary}
    \centering
    \begin{tabular}{c c c c c}
\toprule
\textbf{Degree} & \textbf{Driving-only} & \textbf{Measured-LAI} & \textbf{Coupled (WSO)} & \textbf{Coupled (LAI)} \\
\midrule
1 & 3/5 & 3/10 & 1/10 & 3/10 \\
2 & 6/15 & 9/30 & 7/30 & 7/30 \\
3 & 9/35 & 16/70 & 16/70 & 24/70 \\
4 & 9/70 & 16/140 & 16/140 & 16/140 \\
\bottomrule
\end{tabular}

\end{table}

The RMSE is computed on simulated $S$-trajectories for the held-out test year
in each outer fold, using the model refitted on all outer-training years with
the hyperparameters selected in the inner loop. The largest reduction in RMSE
occurs between $p=1$ and $p=2$. Measured-LAI and Coupled models show a further
smaller improvement at $p=3$, whereas the Driving-only error is similar at
$p=2$ and $p=3$. Increasing the polynomial degree to $p=4$ does not provide a
consistent improvement: it increases RMSE for all three formulations. At $p=3$,
the three library compositions perform similarly, with mean RMSE values around
500 kg ha$^{-1}$. These results indicate that polynomial degrees $p=2$ and
$p=3$ provide sufficient flexibility for predictive performance in this
dataset, whereas increasing the polynomial degree from $p=3$ to $p=4$ adds
complexity without improving generalisation to held-out years.

\begin{figure}[H]
    \centering
    \includegraphics[width=0.9\linewidth]{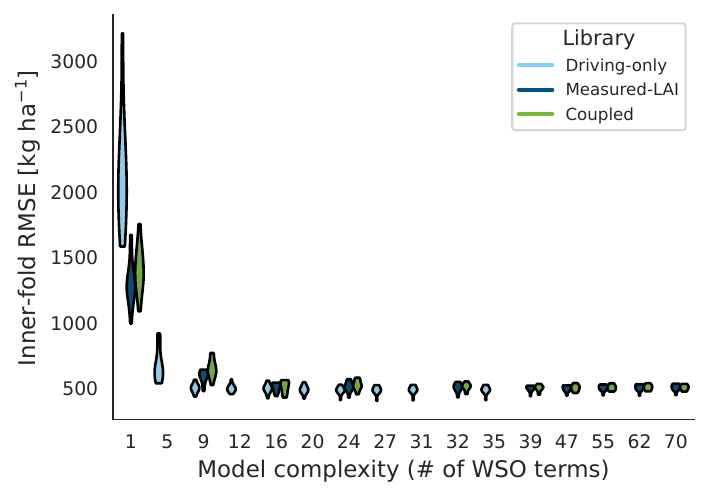}
    \caption{Inner-fold root mean squared error (RMSE) of WSO trajectories as a function of selected WSO model complexity at polynomial degree $p=3$ for the Driving-only, Measured-LAI, and Coupled formulations. For the Coupled formulation, each point at a given WSO complexity corresponds to the LAI complexity with the lowest inner-fold RMSE within that outer fold.}\label{fig:inner_rmse_vs_complexity}
    \figalttext[Validation error as a function of selected WSO model complexity.]{Violin plots of inner-fold WSO RMSE against the number of selected WSO terms for the Driving-only, Measured-LAI, and Coupled formulations at polynomial degree three. RMSE is high and variable for the smallest models, decreases rapidly as the number of selected terms increases, and then remains around a broad plateau of approximately 500~kg~ha$^{-1}$ across larger model complexities.}
\end{figure}

The Coupled library achieves the lowest RMSE at $p=1$, whereas the Driving-only
library is marginally best at $p=2$. This indicates that direct information on
leaf area can be informative for predicting WSO dynamics, but its advantage
depends on the library composition and degree. In the coupled library, $S$
depends on the predicted $\widehat{L}(t)$ rather than the observed $L(t)$, so
errors in $\widehat{L}(t)$ can propagate into $S(t)$.

Using the 1-SE rule substantially reduces the selected model complexity while
retaining similar predictive performance. This is consistent with
Figure~\ref{fig:inner_rmse_vs_complexity}, where validation RMSE often reaches
a broad plateau before the minimum-RMSE complexity is reached. The 1-SE rule
therefore provides a pragmatic accuracy--complexity trade-off and favours
compact models when several candidate complexities perform similarly.
Alternatively, a more formal model-selection criterion, for example
Pareto-front selection~\citep{schmidt2009distilling} or information
criteria~\citep{zhang2023information}, could also be used to select a model
with a good accuracy--complexity trade-off.

We additionally assessed numerical stability and trajectory plausibility of all
held-out simulations. No integration failures or exploding WSO trajectories
occurred, although occasional negative states and non-monotone behaviour were
observed, particularly for coupled LAI at higher polynomial degrees during the
end of the growth window. These issues are discussed in more detail in the
Supplementary Material.

\subsection{Comparison with dynamic baseline models}

Table~\ref{tab:baseline_comparison} compares the SSR-SINDy models with several
dynamic baselines. The simple cultivar-specific mean-trajectory and Gompertz
baselines show substantially larger prediction errors, with mean WSO RMSE
values of $2222.6$ and $2234.1$~kg~ha$^{-1}$, respectively. This indicates that
cultivar-specific average growth patterns alone do not capture the year-to-year
variation in WSO, whereas explicit use of time-varying environmental drivers
substantially improves prediction across held-out years.

\begin{table}[h]
    \centering
    \caption{Comparison of SSR-SINDy with dynamic baseline models. Entries
        report test RMSE of storage-organ dry weight $S$ (mean $\pm$ sample SD,
        kg ha$^{-1}$) across held-out cultivar--year trajectories. Mean-trajectory
        and Gompertz baselines are aligned by temperature integral. For the
        polynomial ODE models, results are shown at polynomial degree $p=3$.
        For the Neural ODE models, results are shown for three independently
        initialised models (with different random seeds).}
    \label{tab:baseline_comparison}
    \begin{tabular}{l l c}
    \toprule
    \textbf{Model}  & \textbf{Library}     & \textbf{WSO RMSE}   \\
    \midrule
    Mean trajectory & Temperature integral & 2222.6 $\pm$ 1197.5 \\
    Gompertz        & Temperature integral & 2234.1 $\pm$ 1183.2 \\
    \midrule
    Ridge ODE       & Driving-only ($p=3$) & 440.8 $\pm$ 270.9   \\
    Ridge ODE       & Measured-LAI ($p=3$) & 517.2 $\pm$ 255.1   \\
    Ridge ODE       & Coupled ($p=3$)      & 546.6 $\pm$ 384.1   \\
    \midrule
    SSR-SINDy       & Driving-only ($p=3$) & 499.4 $\pm$ 230.4   \\
    SSR-SINDy       & Measured-LAI ($p=3$) & 527.8 $\pm$ 337.4   \\
    SSR-SINDy       & Coupled ($p=3$)      & 523.1 $\pm$ 333.9   \\
    \midrule
    Neural ODE      & Driving-only         & 468.7 $\pm$ 220.5   \\
    Neural ODE      & Measured-LAI         & 521.3 $\pm$ 279.9   \\
    Neural ODE      & Coupled              & 544.9 $\pm$ 257.0   \\
    \bottomrule
\end{tabular}

\end{table}

The dense Ridge polynomial ODE baseline performs similarly to SSR-SINDy. At
$p=3$, Ridge yields mean RMSE values of $440.8$, $517.2$, and
$546.6$~kg~ha$^{-1}$ for the Driving-only, Measured-LAI, and Coupled
formulations, respectively, compared with $499.4$, $527.8$, and
$523.1$~kg~ha$^{-1}$ for SSR-SINDy. Thus, sparse term selection does not
provide a uniform improvement in predictive accuracy over the corresponding
dense polynomial ODE models. Its principal contribution is instead explicit
control over term selection and, in many cases, a reduced model structure,
whereas Ridge retains the full candidate library.

The rollout-trained Neural ODE achieves mean RMSE values of $468.7$, $521.3$,
and $544.9$~kg~ha$^{-1}$ for the Driving-only, Measured-LAI, and Coupled
formulations, respectively. It therefore substantially outperforms the simple
trajectory baselines, and is comparable to the SSR-SINDy and Ridge polynomial
ODE models.

Overall, these comparisons suggest that the main predictive gain arises from
representing WSO dynamics as a function of time-varying environmental
conditions. Within this class of models, SSR-SINDy achieves performance
comparable to the dense Ridge-regularised polynomial ODE alternative while
providing an explicit equation-based representation with data-driven term
selection. The Neural ODE offers a flexible continuous-time comparator with
mixed trajectory accuracy relative to the polynomial ODE models: it outperforms
SSR-SINDy for the Driving-only and Measured-LAI formulations, but not for the
Coupled formulation.

\subsection{Stability of identified models}

Strong predictive performance does not necessarily imply that the mathematical
representation of the identified model is uniquely determined by the available
data. Crop-model calibration is generally an ill-posed inverse problem:
correlated environmental variables and limited environmental variation allow
different parameter sets or model structures to describe similar trajectories.
The SINDy formulation is subject to similar issues, since correlated polynomial
candidate functions can provide alternative representations of the same
growth-rate dynamics over the observed environmental conditions.

This raises three distinct questions: whether the same terms are consistently
selected across training sets, whether the GSCs are reproducible once the model
support is fixed, and whether different selected support sets nevertheless
produce similar fitted growth-rate functions under the same environmental
conditions. We refer to these as \textit{selection stability},
\textit{coefficient stability}, and \textit{functional stability},
respectively. The subsequent interpretation of the identified models
(Section~\ref{subsec:interpretability}) is considered in light of these
stability assessments.

\subsubsection{Selection stability}\label{subsubsec:selection_stability}

Selection stability characterises the sensitivity of the selected terms
(referred to as the\textit{support}) to the available training data. We
evaluate it using the 12 outer folds of the nested LOYO-CV procedure, each
obtained after excluding a different year. For each library composition and
polynomial degree, we quantify overlap between the selected terms using the
Jaccard index,
\begin{equation}
    J(\mathcal{S}_i,\mathcal{S}_j)
    =
    \frac{|\mathcal{S}_i\cap\mathcal{S}_j|}
    {|\mathcal{S}_i\cup\mathcal{S}_j|},
\end{equation}
where $\mathcal{S}_i$ and $\mathcal{S}_j$ are the selected supports for outer folds $i$ and $j$. We compute the Jaccard index for all $\binom{12}{2}=66$ pairwise
comparisons by their median and
interquartile range (IQR). The Jaccard index ranges from 0 (no overlap) to 1
(perfect overlap), with higher values indicating greater selection stability.
Table~\ref{tab:stability_combined} shows that support overlap varies with
candidate-library composition and polynomial degree. For the driving-only
library, for example, the median Jaccard similarity decreases from $1.0$ at
$p=1$ to $0.636$ at $p=4$. This is expected since the highly correlated
environmental variables and polynomial terms provide alternative
representations of similar dynamics. Thus, richer libraries do not necessarily
lead to one unique polynomial decomposition of the growth rate. Importantly, a
unique support is not required for high predictive performance. When several compact supports
describe the observed dynamics equally well, the model-selection criterion can
select one suitable representation out of these supports for subsequent cultivar-specific coefficient
estimation. Moreover, patterns of support non-identifiability can indicate
which additional environmental conditions or measurements may help distinguish
between alternative representations, helping effective experimental design.

\subsubsection{Coefficient stability}
\label{subsubsec:coefficient_stability}

\begin{figure}[H]
    \centering
    \includegraphics[width=1.0\linewidth]{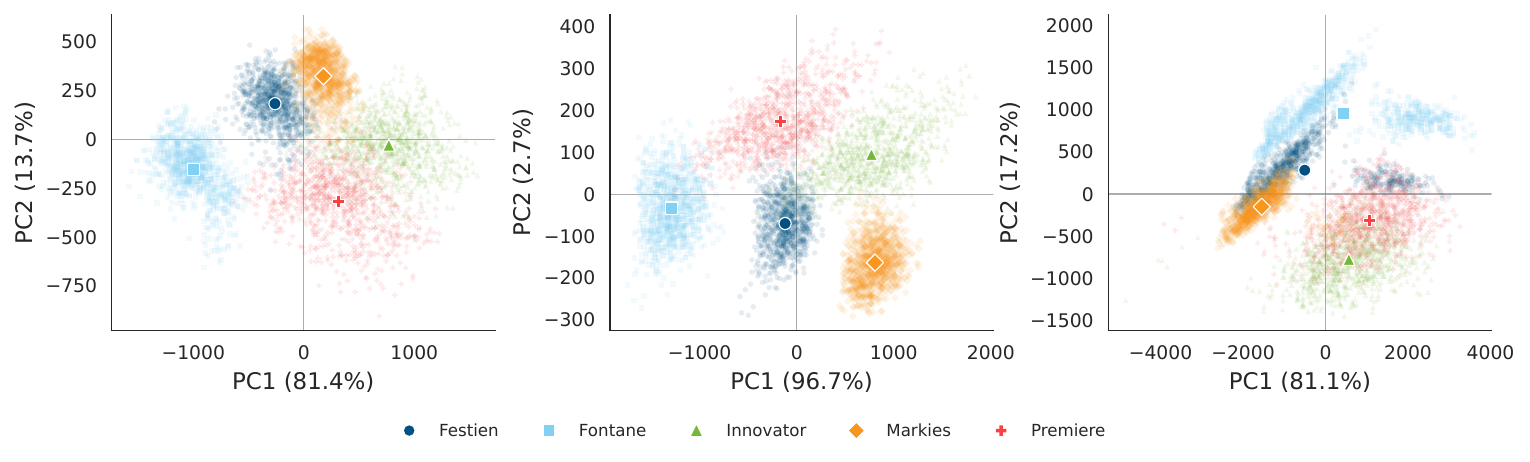}
    \caption{Cultivar-specific variation in contribution-scaled genotype-specific coefficient (GSC) vectors under calibration-environment resampling for the Driving-only SINDy formulation at $p=3$. Each panel corresponds to a different outer cross-validation fold, with 1986, 1989, or 1991 held out. Within each outer fold, the SINDy support is fixed to that selected from the corresponding outer-training data. The cultivar-specific coefficients are then re-estimated by OLS for all $\binom{12}{6}=924$ subsets of six calibration environments. Each point represents the resulting coefficient vector for one cultivar and one six-environment subset; colours and marker symbols distinguish cultivars, and enlarged markers indicate cultivar-wise means. PCA is performed separately for each outer-fold support.}\label{fig:coefficient_pca}
    \figalttext[
        PCA of cultivar-specific coefficient variation under calibration-environment resampling.
    ]{
        Three PCA scatter plots show contribution-scaled GSC vectors for the
        Driving-only model at polynomial degree three for outer folds holding out
        1986, 1989, and 1991. Each point represents one of the 924 six-environment
        fixed-support coefficient refits, with cultivars distinguished by colour and
        cultivar-specific marker symbols; enlarged markers indicate cultivar-wise
        mean vectors. Cultivar-specific clusters are
        visible in all three panels, although the degree of separation and
        within-cultivar dispersion differs among outer folds. PCA is performed
        independently for each outer-fold support, so the principal-component axes are
        not directly comparable across panels.
    }
\end{figure}

Next, we assess whether cultivar-specific GSC differences are stable to changes
in the environments used for calibration. Because the GSCs are conditional on
the selected support, we separate coefficient stability from support selection
by fixing the support within each outer fold. For outer fold $r$, we use the
support $\mathcal{S}_r$ selected from the 11 outer-training environments and
refit the GSCs on all $B=\binom{12}{6}=924$ subsets of six environments drawn
from the full set of 12 environments. The support is held fixed across all
these 924 subsets, so that the analysis isolates coefficient variation induced
by the environments used for coefficient estimation. This resampling analysis
is used only as a post-hoc stability diagnostic and does not enter model
selection or held-out-year prediction.

For each selected term $j\in\mathcal{S}_r$, let $\xi_{j,S,r}^{(g,b)}$ denote
its coefficient for cultivar $g$ and calibration subset $b$. We decompose each
coefficient as follows:
\begin{equation}
    \xi_{j,S,r}^{(g,b)}
    =
    \mu_{j,r}
    +
    a_{j,r}^{(g)}
    +
    c_{j,r}^{(b)}
    +
    (ac)_{j,r}^{(g,b)},
\end{equation}
where $a_{j,r}^{(g)}$ represents persistent cultivar differences,
$c_{j,r}^{(b)}$ represents shifts common to all cultivars for a given
calibration subset, and $(ac)_{j,r}^{(g,b)}$ represents cultivar-specific
sensitivity to the calibration subset, here coinciding with the residual term. For each of these terms, we estimate the variances $\sigma^2_{\mathrm{cult}}$,
$\sigma^2_{\mathrm{cult}\times\mathrm{sub}}$, and $\sigma^2_{\mathrm{sub}}$. Then, we quantify coefficient stability using the single-measurement consistency
intraclass correlation coefficient, $\mathrm{ICC}(C,1)$
\citep{mcgraw1996forming},
\begin{equation}
    \mathrm{ICC}(C,1)
    =
    \frac{\sigma^2_{\mathrm{cult}}}
    {\sigma^2_{\mathrm{cult}}
    +\sigma^2_{\mathrm{cult}\times\mathrm{sub}}}.
\end{equation}
This measures the extent to which relative cultivar differences in a
coefficient are preserved across calibration subsets. A value close to 1
indicates that persistent cultivar differences dominate over
cultivar-specific sensitivity to the calibration subset, whereas a value close
to 0 indicates poor preservation of cultivar differences across subsets.

Table~\ref{tab:coefficient_stability} reports the median consistency ICC over
the selected terms of each outer-fold support, summarised across the 12 outer
folds. At $p=2$, median ICC values ranged from $0.853$ to $0.943$ across the
three WSO formulations. The effect of increasing polynomial degree depended on
the formulation. For Driving-only WSO, the median ICC increased from $0.853$ at
$p=2$ to $0.925$ at $p=3$ and $0.968$ at $p=4$, although variability across
outer-fold supports increased at $p=4$. For Measured-LAI and Coupled WSO, the
median ICC decreased to $0.658$ at $p=3$ before increasing to $0.800$ at $p=4$.
Coupled LAI showed consistently high median ICC values across degrees, ranging
from $0.845$ to $0.983$.

\begin{table}[h]
    \centering
    \caption{Consistency intraclass correlation coefficient,
        $\mathrm{ICC}(C,1)$, of cultivar-specific coefficients under resampling of
        six-environment calibration subsets. For each outer-fold support, the median
        over selected terms is computed; entries show the median [$Q_1$--$Q_3$]
        across the 12 outer folds. Higher values indicate greater preservation of
        relative cultivar differences across calibration subsets.}
    \label{tab:coefficient_stability}
    \resizebox{0.85\linewidth}{!}{
        \begin{tabular}{lcccc}
\toprule
\textbf{Degree} & \textbf{Driving-only (WSO)} & \textbf{Measured-LAI (WSO)} & \textbf{Coupled (WSO)} & \textbf{Coupled (LAI)} \\
\midrule
$p=1$ & 0.931 [0.931--0.931] & 0.953 [0.920--0.996] & 0.996 [0.923--0.996] & 0.983 [0.966--0.983] \\
$p=2$ & 0.853 [0.853--0.855] & 0.881 [0.784--0.943] & 0.943 [0.943--0.943] & 0.908 [0.888--0.916] \\
$p=3$ & 0.925 [0.919--0.952] & 0.658 [0.658--0.823] & 0.658 [0.654--0.658] & 0.870 [0.769--0.882] \\
$p=4$ & 0.968 [0.640--0.968] & 0.800 [0.639--0.860] & 0.800 [0.639--0.860] & 0.845 [0.827--0.858] \\
\bottomrule
\end{tabular}

    }
\end{table}

Once the support is fixed, the GSCs therefore generally retain reproducible
cultivar-specific differences across calibration subsets, although the degree
of stability depends on the library formulation and polynomial degree. The
lower ICC values for Measured-LAI and Coupled WSO at $p=3$ indicate greater
sensitivity of some coefficients to calibration-set composition, whereas the
Driving-only and Coupled LAI formulations show stronger preservation of
cultivar differences across most evaluated degrees. Because the calibration
subsets overlap strongly and only five cultivars are available, the ICCs are
interpreted purely descriptively. Nevertheless, the results indicate that
cultivar-specific differences in the GSCs can remain reproducible across
substantial changes in calibration environments, which supports considering
them as cultivar-specific descriptors and potential candidate traits for future
genomic analysis.

To complement the term-wise ICC analysis, we visualise the joint variation of
the GSCs for the Driving-only $p=3$ model, by decomposing the GSC vectors.
Before PCA, each GSC is multiplied by the root-mean-square value of its
corresponding library term over the full 12-year design for that cultivar. This
places coefficients on a common scale, so that the PCA reflects differences in
their contributions to the growth rate rather than differences in coefficient
scale. PCA is performed separately within each outer fold because the selected
support can differ between folds, using all 924 six-environment refits and
colouring points by cultivar. Figure~\ref{fig:coefficient_pca} shows that
cultivar-specific clusters remain visible across the resampled fits, although
the degree of separation varies among outer-fold supports.

\subsubsection{Functional stability}\label{subsubsec:functional_stability}
Finally, selected supports and GSCs may vary across outer folds while producing
similar fitted growth-rate functions, due to the inherent correlation between
the selected terms and the environmental conditions. To assess this functional
stability, we evaluate the inner-fold models from each outer fold on the
environmental trajectory of the corresponding held-out year, for a total of 132
evaluations per library-degree combination. For the Measured-LAI formulation,
observed WOFOST $L(t)$ is supplied as usual. For the Coupled formulation, we
likewise use observed $L(t)$ when evaluating the WSO rate function, thereby
isolating variation in the fitted WSO response from errors in the predicted LAI
trajectory.

Let $\widehat R_{S,h}(t_i)$ denote the WSO growth rate predicted by inner-fold
model $h=1,\ldots,11$ at time $t_i$ in the held-out outer year, and
$\overline{R}_S(t_i) = \tfrac{1}{11}\sum_h \widehat R_{S,h}(t_i)$ their mean.
We summarise relative functional variability across inner folds as
\begin{equation}
    V_S =
    \frac{
        \sqrt{\tfrac{1}{n}\sum_{i=1}^{n}
            \operatorname{SD}_h^2\!\left[\widehat R_{S,h}(t_i)\right]}
    }{
        \operatorname{SD}_i\!\left[\overline R_S(t_i)\right]
    },
\end{equation}
where $n$ is the number of time points in the held-out year and
$\operatorname{SD}_h[\cdot]$ and $\operatorname{SD}_i[\cdot]$ denote the
standard deviation across inner folds and time points, respectively. Thus,
$V_S$ measures the typical disagreement among models relative to the temporal
variation in the mean predicted WSO growth rate. Smaller values indicate
greater functional stability under the evaluated environmental conditions. As a
separate measure of functional accuracy, we additionally report the $R^2$
between the WSO rate predicted by the final outer-fold model and the
corresponding WOFOST daily storage-organ growth rate.

\begin{table}[h]
    \centering
    \caption{Selection stability, functional stability, and direct growth-rate
        accuracy of identified WSO models by library composition and polynomial
        degree. Support Jaccard is the median [$Q_1$--$Q_3$] pairwise similarity
        of the shared WSO support across the 12 outer folds
        ($\binom{12}{2}=66$ pairs). $V_S$ quantifies between-model variability of
        the fitted WSO growth-rate function among the 11 inner-fold models within
        an outer fold, evaluated under matched held-out-year conditions; entries
        show the median [$Q_1$--$Q_3$] across outer folds and cultivars, where
        lower values indicate greater functional stability. Growth-rate $R^2$ is
        the median direct daily $R^2$ of the final outer-fold WSO model against
        WOFOST storage-organ growth rates in the held-out year, using observed
        WOFOST covariates. For Measured-LAI and Coupled WSO, these RHS
        diagnostics use observed LAI, thereby isolating the fitted WSO rate
        function from LAI rollout error. Bold entries mark the best value within each library
        composition for the respective metric.}\label{tab:stability_combined}
    \resizebox{0.85\linewidth}{!}{%
        \begin{tabular}{c c c c c}
\toprule
\textbf{Library} & \textbf{Degree} & \textbf{Support Jaccard} & \textbf{$V_S$} & \textbf{Growth-rate $R^2$} \\
\midrule
Driving-only & 1 & \textbf{1.000 [0.667--1.000]} & 0.207 [0.171--0.309] & -0.019 \\
 & 2 & 1.000 [0.714--1.000] & \textbf{0.063 [0.051--0.078]} & 0.746 \\
 & 3 & 0.556 [0.286--0.800] & 0.088 [0.065--0.145] & 0.782 \\
 & 4 & 0.636 [0.500--1.000] & 0.113 [0.102--0.150] & \textbf{0.802} \\
\addlinespace
Measured-LAI & 1 & 0.333 [0.200--0.400] & 0.204 [0.150--0.254] & 0.514 \\
 & 2 & 0.636 [0.500--0.786] & \textbf{0.076 [0.055--0.096]} & 0.777 \\
 & 3 & \textbf{0.684 [0.562--0.882]} & 0.081 [0.069--0.108] & 0.820 \\
 & 4 & 0.455 [0.371--0.600] & 0.135 [0.117--0.161] & \textbf{0.824} \\
\addlinespace
Coupled & 1 & 0.250 [0.143--1.000] & 0.162 [0.130--0.228] & 0.510 \\
 & 2 & \textbf{1.000 [0.556--1.000]} & 0.097 [0.063--0.146] & 0.782 \\
 & 3 & 0.882 [0.684--1.000] & \textbf{0.078 [0.067--0.086]} & \textbf{0.845} \\
 & 4 & 0.455 [0.371--0.600] & 0.135 [0.117--0.161] & 0.824 \\
\bottomrule
\end{tabular}

    }
\end{table}

\subsubsection{Implications for identifiability and interpretation}
\label{subsubsec:identifiability_implications}

Taken together, the stability analyses show that different aspects of the
identified models have different levels of stability. The exact polynomial
support is not uniquely determined by the data, particularly for richer
libraries, consistent with correlations among environmental variables and
polynomial candidate functions. The selected support should therefore be
regarded as one suitable sparse representation rather than as a uniquely
identified physiological structure. Representative Driving-only $p=3$ equations
from different outer folds, together with their unscaled cultivar-specific
coefficients and physical units, are provided in Supplementary Material, and further illustrate this
variability in the selected sparse representation.

Conditional on a chosen support, however, the GSCs generally preserve
cultivar-specific differences across calibration subsets. They can therefore be
considered reproducible cultivar-level quantitative descriptors and potential
candidate traits for future genomic analysis, although their interpretation
remains conditional on the selected support and they should not be equated
directly with physiological parameters.

Finally, different selected supports can nevertheless produce similar fitted
growth-rate responses under the environmental conditions represented in the
data. Physiological interpretation is therefore most appropriately based on the
resulting functional responses rather than on individual polynomial terms.

\subsection{Interpretability of the identified growth-rate models}\label{subsec:interpretability}

\begin{figure}[H]
    \centering
    \includegraphics[width=\linewidth]{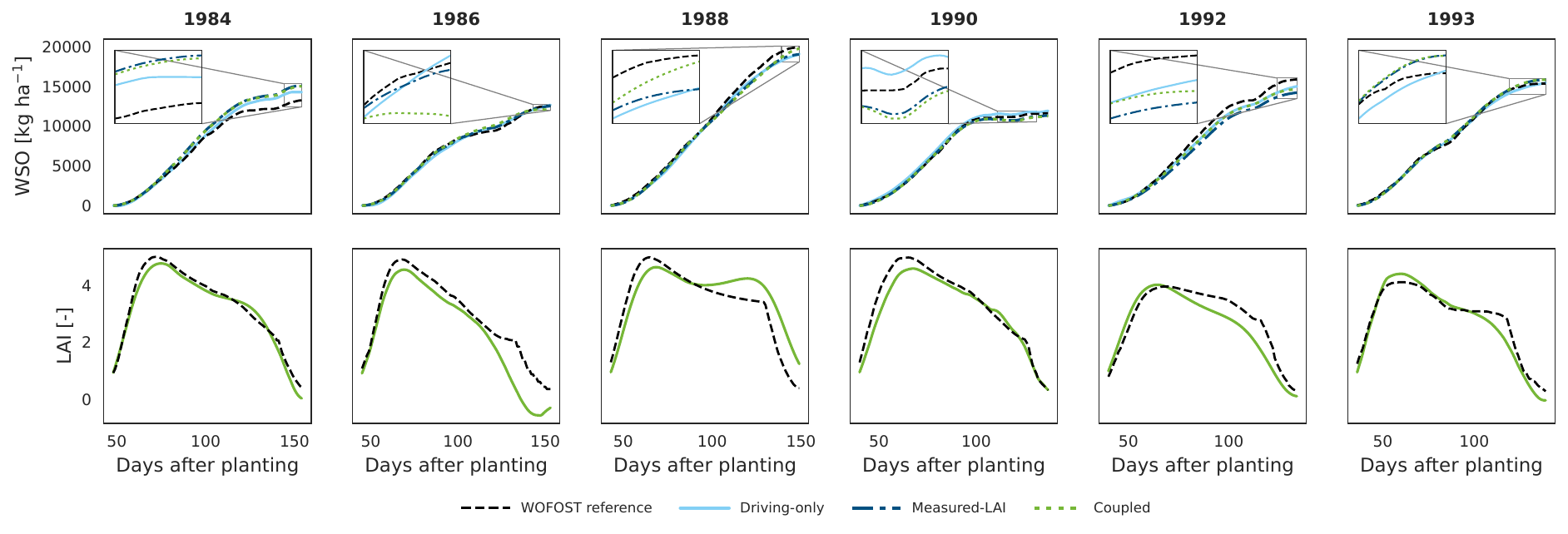}
    \caption{Selected WSO and LAI trajectories for cultivar Premiere in six held-out years at polynomial degree $p=3$. The years comprise the three environments with the lowest and three with the highest mean WSO trajectory RMSE across the Driving-only, Measured-LAI, and Coupled SINDy models. Upper panels show WSO predictions from the three formulations and the WOFOST reference trajectory. Inset plots highlight the differences among the different libraries. Lower panels show the Coupled-model LAI prediction and its WOFOST reference trajectory.}\label{fig:representative_trajectories}
    \figalttext[
        Predicted and reference crop-growth trajectories in six held-out years.
    ]{
        Six year panels compare predicted and reference storage-organ growth.
        Agreement is strongest in the low-error years and weaker near the end of
        the season in the high-error years. Coupled leaf-area trajectories broadly
        follow reference development and senescence.
    }
\end{figure}

Before examining the functional responses of the identified models, we first
illustrate their dynamical behaviour in held-out environments.
Figure~\ref{fig:representative_trajectories} shows predicted and reference
trajectories for Premiere in six selected held-out years. All three library
compositions reproduce the overall seasonal accumulation of WSO, although
differences become more apparent near the end of the growing period and vary
among years. The Coupled model also reproduces the main seasonal development
and senescence pattern of LAI.

\begin{figure}[H]
    \centering
    \includegraphics[width=\linewidth]{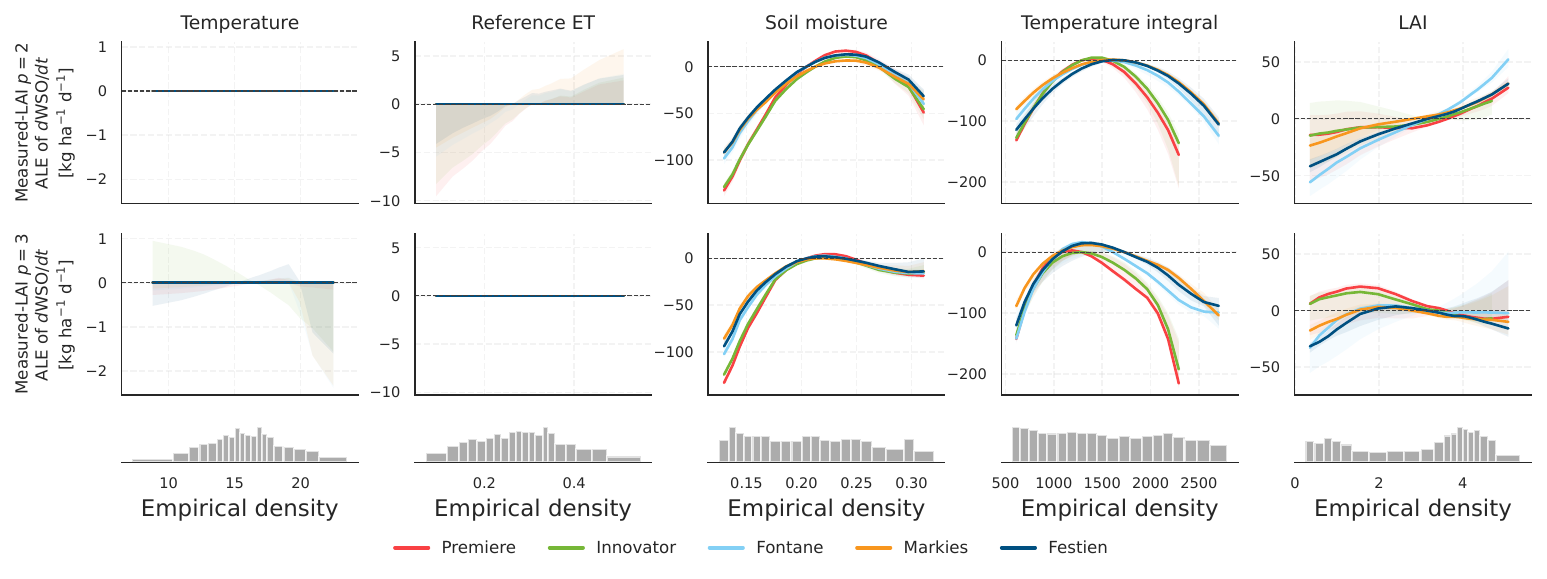}
    \caption{Cultivar-specific first-order accumulated local effect (ALE) curves
        for the Measured-LAI WSO growth-rate models. Rows show different
        polynomial degrees
        $p=2$, $p=3$, and columns show temperature, reference
        evapotranspiration, soil moisture, temperature integral, and LAI.
        Cultivar-specific curves show the median ALE across the 12 outer-fold
        models, with shaded regions indicating the corresponding 10--90\% fold
        range. Histograms below each column show the pooled empirical distribution
        of the corresponding predictor over active-growth observations.}
    \label{fig:ale_responses}
    \figalttext[
        Cultivar-specific accumulated local effects across five predictors and two polynomial degrees.
    ]{
        Ten accumulated local effect plots, with five predictor-density histograms,
        show broadly similar response shapes across polynomial degrees two and three. Cultivar differences are clearest
        for temperature integral, where Premiere and Innovator decline earlier
        than the other cultivars.
    }
\end{figure}

The stability analyses above show that model support can vary across training
sets, while GSCs can remain reproducible conditional on a fixed support and
fitted growth-rate functions can remain similar across alternative supports. We
therefore use functional responses to provide a support-independent
interpretation of the identified models.

To characterise these responses, we use accumulated local effects
(ALE)~\citep{apley2020visualizing}, which summarise the average local effect of
a predictor on the fitted WSO growth-rate function over the empirical data
distribution. For a variable $v_j\in\{L,u_1,\ldots,u_q\}$, the first-order ALE
is defined as
\begin{equation}
    \mathrm{ALE}_{j}(v)
    =
    \int_{v_{j,0}}^{v}
    \mathbb{E}\!\left[
        \left.
        \frac{\partial R_S(L,\mathbf{u},\boldsymbol{\xi}_S)} {\partial v_j} \right| v_j=z \right]\,dz - C_j ,
\end{equation}
where $C_j$ anchors the ALE at the empirical median of $v_j$. The slope
of the ALE curve therefore describes the average local sensitivity of the
fitted growth rate to that predictor over the observed data distribution. ALE
curves are computed separately for each outer-fold model using a common
cultivar-specific empirical predictor distribution and are then summarised
across folds by their median and 10--90\% range.

Figure~\ref{fig:ale_responses} shows that the main environmental-response
patterns are generally consistent across outer folds despite variation in the
selected supports and coefficients. The dominant response shapes are already
apparent at $p=2$ and remain broadly similar at $p=3$, suggesting that
increasing polynomial degree beyond $p=3$ introduces relatively little
additional change in the qualitative behaviour of the fitted environmental
responses.

Clear cultivar differences occur in the response to the temperature integral.
For both degrees $p=2$ and $p=3$, Premiere and Innovator show an earlier
decline in the ALE response than the other three cultivars. Interestingly,
these two cultivars also have the lowest WOFOST thermal-time requirement from
emergence to tuber initiation (TSUM1): 164~$^\circ$Cd for Premiere and
187~$^\circ$Cd for Innovator, compared with 222--280~$^\circ$Cd for Fontane,
Markies, and Festien. Although the timing of the ALE decline should not be
interpreted as a direct estimate of TSUM1, this correspondence suggests that
cultivar-specific phenological differences encoded in the underlying WOFOST
parameterisation are captured by the fitted functional responses.

Overall, the ALE analysis supports the conclusion from the stability analysis
that the fitted growth-rate functions are more stable than the selected
polynomial representations: different polynomial representations produce
similar fitted environmental-response functions over the conditions represented
in the data. Functional response shapes, sensitivities, and developmental
timing may therefore provide more support-independent measures for comparing
cultivars than individual SINDy coefficients.

\subsection{Yield prediction}
\begin{figure}[H]
    \centering
    \includegraphics[width=1.0\linewidth]{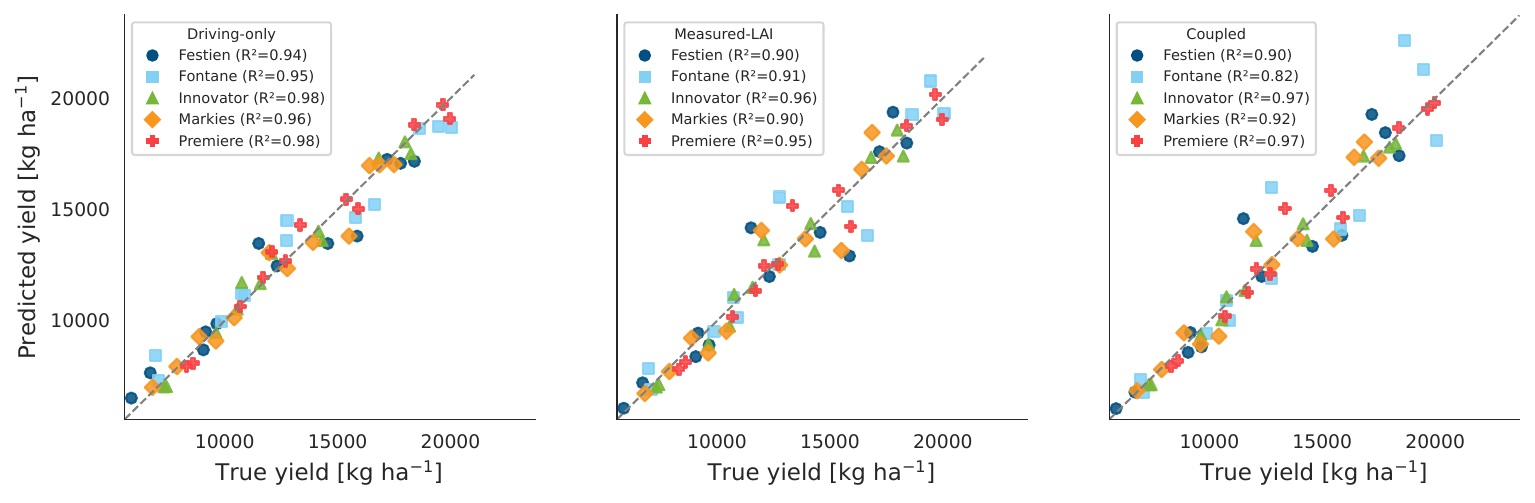}
    \caption{True versus predicted yield in the outer test folds of the nested leave-one-year-out cross-validation procedure. From left to right, panels show the Driving-only, Measured-LAI, and Coupled formulations at polynomial degree $p=3$. Predicted yield is the maximum predicted WSO over the predicted active-growth interval, whereas reference yield is the maximum WSO over the full corresponding WOFOST trajectory. Cultivars are distinguished by colour and marker shape, and the diagonal line denotes perfect agreement.}\label{fig:true_vs_predicted_yield}
    \figalttext[Predicted versus WOFOST reference yield for the three SINDy formulations.]{Three scatter plots compare predicted yield with WOFOST reference yield for the Driving-only, Measured-LAI, and Coupled formulations. Points represent five cultivars and are distinguished by both colour and marker shape. Across all three formulations, most points lie close to the one-to-one line over the full yield range, with some deviations at higher yields.}
\end{figure}

We next evaluate end-of-season yield prediction, defined as the maximum of the
predicted storage-organ biomass $S(t)$ over the predicted growth window.
Results using the endpoint of the predicted growth window, $S(\hat{t}_1^{S})$,
were very similar (Supplementary Material).
Figure~\ref{fig:true_vs_predicted_yield} compares true and predicted yields for
the three SINDy library compositions at polynomial degree $p=3$, using only
outer-fold test predictions.

\begin{table}[h]
    \centering
    \caption{Held-out yield-prediction performance across the 12 outer folds.
        Predicted yield is defined as the maximum predicted WSO over the
        predicted growth window and is evaluated against maximum WSO from the full
        WOFOST trajectory. Overall metrics are computed across all held-out 60
        cultivar--year predictions. Neural ODE entries pool three independently
        initialised models (with different random seed).
        Year-centred metrics remove the held-out-year mean yield, whereas G$\times$E metrics remove both year and
        cultivar main effects. Rank correlation is the mean within-year Spearman
        correlation across cultivars. RMSE is in kg ha$^{-1}$. SSR-SINDy and Ridge polynomial ODE
        models are shown for $p=3$.}
    \label{tab:yield_baseline_comparison}
    \resizebox{0.85\linewidth}{!}{
        \begin{tabular}{lcccccc}
\toprule
\textbf{Model} & \textbf{RMSE} & \textbf{$R^2$} & \textbf{Year-centred $R^2$} & \textbf{Rank $\rho$} & \textbf{G$\times$E $R^2$} & \textbf{G$\times$E $r$} \\
\midrule
\multicolumn{7}{l}{\textit{Trajectory references}} \\
Mean trajectory & 4292 & -0.164 & 0.541 & 0.658 & -0.175 & -0.877 \\
Gompertz & 4313 & -0.176 & 0.517 & 0.650 & -0.204 & -0.611 \\
\addlinespace
\multicolumn{7}{l}{\textit{Dense polynomial ODE}} \\
Ridge, Driving-only & 881 & 0.951 & 0.637 & 0.750 & 0.077 & 0.528 \\
Ridge, Measured-LAI & 940 & 0.944 & 0.762 & 0.808 & 0.390 & 0.682 \\
Ridge, Coupled & 1111 & 0.922 & 0.415 & 0.650 & -0.447 & 0.533 \\
\addlinespace
\multicolumn{7}{l}{\textit{Sparse polynomial ODE}} \\
SSR-SINDy, Driving-only & 800 & 0.960 & 0.738 & 0.767 & 0.341 & 0.621 \\
SSR-SINDy, Measured-LAI & 1089 & 0.925 & 0.793 & 0.783 & 0.474 & 0.703 \\
SSR-SINDy, Coupled & 1195 & 0.910 & 0.418 & 0.792 & -0.474 & 0.546 \\
\addlinespace
\multicolumn{7}{l}{\textit{Neural ODE}} \\
Neural ODE, Driving-only & 882 & 0.951 & 0.508 & 0.703 & -0.144 & 0.461 \\
Neural ODE, Measured-LAI & 1048 & 0.931 & 0.377 & 0.695 & -0.567 & 0.415 \\
Neural ODE, Coupled & 1051 & 0.930 & 0.266 & 0.664 & -0.858 & 0.387 \\
\addlinespace
\multicolumn{7}{l}{\textit{Static yield models}} \\
Cultivar training-year mean & 4285 & -0.160 & 0.537 & 0.633 & -0.190 & -1.000 \\
Ridge seasonal & 3214 & 0.347 & -4.383 & 0.150 & -12.073 & -0.220 \\
Ridge seasonal + LAI & 2075 & 0.728 & -0.866 & 0.617 & -3.679 & 0.250 \\
\bottomrule
\end{tabular}

    }
\end{table}

Across all three SINDy formulations, predicted yields closely track the
reference values, with overall $R^2>0.9$. Because overall yield variation is
dominated by differences among years, however, these metrics mainly reflect the
ability to reproduce environmental variation. We therefore additionally
evaluate year-centred yield variation, within-year cultivar ranking, and the
residual G$\times$E component. By removing the held-out-year mean, year-centred
$R^2$ focuses on within-year cultivar differences and is analogous to the
squared prediction ability commonly reported for genomic prediction under
G$\times$E.

The ability to reproduce cultivar-specific differences varies more strongly
among formulations. Among the SINDy models, the Measured-LAI formulation gives
the highest year-centred $R^2$ ($0.793$) and G$\times$E $R^2$ ($0.474$),
whereas the Coupled formulation gives the highest mean within-year rank
correlation ($\rho=0.792$). The Driving-only and Coupled formulations give
strong overall $R^2$ values ($0.960$ and $0.910$, respectively), but lower
year-centred and G$\times$E $R^2$ values.

The environment-driven dynamic models substantially outperform the static yield
baselines in overall prediction. They also generally preserve more of the
within-year and G$\times$E variation, although performance differs among
formulations. In particular, the Measured-LAI SINDy and dense ODE-based models
retain considerably more G$\times$E structure than the static regression
baselines. For the baselines considered here, this suggests that modelling the
longitudinal growth process may retain cultivar-specific environmental
responses that are not captured by the seasonal-summary baselines considered
here.

For the Neural ODE, overall yield prediction is worse than SSR-SINDy for the
Driving-only formulation but better for the Measured-LAI and Coupled
formulations. However, it does not preserve within-year cultivar ranking or
G$\times$E variation as well.

\subsection{Sensitivity to noise and sampling density}

\begin{figure}[H]
    \centering
    \includegraphics[width=1.0\linewidth]{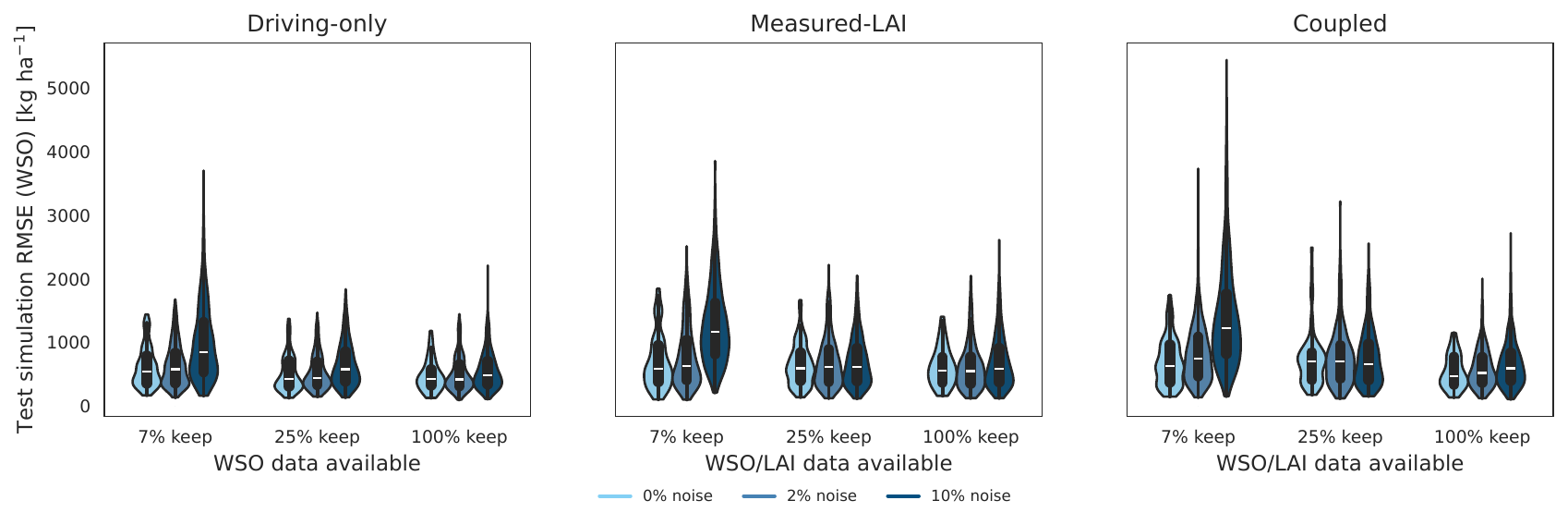}
    \caption{Held-out WSO trajectory root mean squared error (RMSE) under observation noise and data sparsity for the Driving-only, Measured-LAI, and Coupled formulations at polynomial degree $p=2$. Violin plots show held-out cultivar--year RMSE values for retained data fractions of 7\%, 25\%, and 100\%; within each fraction, noise levels of 0\%, 2\%, and 10\% are shown from left to right. Savitzky--Golay polynomial order was fixed at 2, while smoothing-window length and SSR model complexity were selected within the inner leave-one-year-out validation loop. The clean, complete-observation condition is deterministic and contains one realisation; all other conditions contain 10 independent observation-process realisations.}\label{fig:noise_sparsity_violin}
    \figalttext[Prediction error across observation-noise levels and sampling densities.]{Three panels show WSO trajectory RMSE for the Driving-only, Measured-LAI, and Coupled formulations under increasing observation noise and decreasing data availability. Across formulations, RMSE generally increases with noise and is highest when only 7\% of the data are retained, particularly at 10\% noise. Retaining more observations generally reduces the effect of observation noise.}
\end{figure}

For realistic applications, it is important that the identified models can
handle noisy and sparse data. To assess the robustness of the SINDy method, we
add relative Gaussian noise to the trajectories of $S$ and $L$, with noise
levels $\eta \in \{0.0, 0.02, 0.1\}$. Each observation $S(t_j)$ is replaced by
\begin{equation*}
    \hat{S}(t_j) = S(t_j) + \varepsilon(t_j), \qquad \varepsilon(t_j) \sim \mathcal{N}(0, (\eta\, S(t_j))^2),
\end{equation*}
and similarly for $L$, for all observation times $t_j$. These levels represent
three practical regimes: an idealized noise-free case ($\eta=0$), mild
measurement noise ($\eta=0.02$), and a high-noise scenario ($\eta=0.1$) for
derivative estimation. The clean, complete-observation condition is deterministic and was evaluated once. All other conditions were repeated for 10 independently generated observation-process realisations (noise and, where applicable, subsampling).

To simulate data sparsity, we retain only a fraction of the simulated daily
data, subsampled at regular intervals. We consider percentages $\rho \in \{7\%,
    25\%, 100\%\}$, corresponding to approximately 8, 25, and 100 observations per
growing season, respectively (depending on season length). These settings mimic
typical monitoring frequencies from sparse field campaigns (weekly-like), to
moderate revisit schedules, to dense daily measurements.

Derivatives were estimated using Savitzky--Golay filtering with polynomial
order fixed at $q=2$. The smoothing-window length was treated as a
preprocessing hyperparameter, with candidate values
$w\in\{3,5,7,9,11,13,15,17\}$ observations, restricted to feasible values for
the available trajectory length. Within each outer training fold, SSR
complexity was selected separately for each smoothing window using the 1-SE
rule; the window with the lowest resulting inner-LOYO WSO trajectory RMSE was
then retained. The selected configuration was refitted using the complete
outer-training data before evaluation on the outer held-out year.

Figure~\ref{fig:noise_sparsity_violin} shows the resulting test RMSE values for
fixed polynomial degree $p=2$ for the different library formulations. As
expected, increasing noise leads to higher RMSE due to reduced accuracy of
derivative estimation. For noise-free and low-noise data ($\eta = 0$ and $\eta
    = 0.02$), increased data availability generally improves performance.

For the highest noise level ($\eta = 0.1$), the strongest degradation occurs
when high noise is combined with sparse sampling. Retaining 25\% or 100\% of
the observations substantially reduces RMSE relative to retaining only 7\%.
Under these noisy conditions, the selected Savitzky--Golay windows generally
increase with sampling density and noise level, indicating that stronger
smoothing is preferred when sufficient observations are available (see
Supplementary Material).

\section{Discussion}\label{sec:discussion}
In this work, we demonstrated that reduced ODE models describing crop-growth
dynamics can be identified directly from longitudinal data using the SINDy
framework. Although demonstrated only on an \textit{in silico} dataset
generated with WOFOST, the framework appears promising for learning explicit
crop-growth models from longitudinal phenotyping data, given sufficiently dense
and accurate observations. The identified ODEs reproduced growth trajectories
in held-out years within the environmental domain represented by the dataset
and captured a large part of the associated variation in final yield.
Importantly, the identified GSCs also retained reproducible cultivar-specific
differences across substantial changes in the calibration environments.
Comparison with alternative dynamic models showed, however, that much of the
predictive gain arises from representing crop growth as an environmentally
driven dynamical system, while SINDy provides an explicit and potentially
reduced representation of those dynamics.

Looking back at the goals outlined in the introduction, we can summarise our
findings as follows:
\begin{itemize}

      \item \textbf{Goal 1: Describing crop growth dynamics under environmental variation.}
            The SINDy-identified ODEs achieve low test RMSE in held-out years within
            the range of environmental conditions represented in the training data.
            Polynomial degrees $p=2$--$3$ were sufficient to obtain accurate WSO
            trajectory predictions, whereas increasing the degree to $p=4$ did not improve predictive
            performance in held-out years. The models also reproduced much of the variation
            in final yield. In particular, the measured-LAI formulation retained
            substantial within-year cultivar differences and G$\times$E-related
            variation. These results demonstrate feasibility in a controlled setting,
            but remain to be validated on real-world phenotyping data.

      \item \textbf{Goal 2: Explicit control of model complexity.}

            The SINDy framework~\citep{brunton2016discovering} is designed to promote
            sparse models by selecting a subset of candidate terms. However,
            prediction-based cross-validation did not always select highly sparse models,
            particularly for richer candidate libraries, and models with different levels
            of complexity often achieved similar predictive performance.
            Figure~\ref{fig:inner_rmse_vs_complexity} shows that prediction error
            frequently reaches a plateau before the largest selected model sizes. The
            1-standard-error rule therefore provides a practical way to favour more compact
            models while retaining similar predictive performance.

      \item \textbf{Goal 3: Compatibility with genetic modelling.}
            The identified models are linear in their parameters, which makes them
            naturally compatible with statistical approaches for jointly modelling
            cultivar-specific coefficients~\citep{technow2015integrating,poudel2023integrating}.
            The stability analysis further shows that cultivar-specific differences in the
            GSCs are generally reproducible across calibration environments once a model
            representation has been selected. This supports their use as cultivar-level
            quantitative responses in subsequent genetic or genomic modelling. However, this extension remains to be demonstrated in practice in future work.
      \item \textbf{Goal 4: Amenability to stochastic and statistical analysis.}
            Similarly, the linear-in-parameters structure of the identified ODEs
            provides a convenient basis for statistical approaches to parameter
            estimation and uncertainty quantification. Structured regularization~\citep{argyriou2006multi,obozinski2010joint},
            as well as Bayesian extensions~\citep{boninsegna2018sparse,fung2024rapid} could
            be used to account for uncertainty in coefficients, model structure, and
            derived functional responses. Such statistical extensions remain an
            important direction for future work.
\end{itemize}

Our results also distinguish three aspects of model performance. The varying
supports indicate limited identifiability of the exact polynomial
representation. However, the coefficient-stability analysis shows that
cultivar-specific differences in the GSCs are, once the support is selected,
generally consistent under different calibration environments. The GSCs
therefore provide compact cultivar-level descriptors, whereas the fitted
functions provide a complementary basis for interpreting cultivar-specific
responses to environmental conditions. Finally, the held-out-year results
demonstrate predictive generalisation within the environmental domain
represented by the data.

Despite these promising results, several limitations remain. First, the
analysis is based entirely on synthetic data generated by WOFOST for one crop,
one location, and five cultivars. The leave-one-year-out validation therefore
tests prediction in held-out years within the environmental domain represented
by this simulation experiment; it does not establish generalisation to new
locations, climates, crops, management practices, soil conditions, or real
phenotyping data. Application to field data will also introduce measurement
noise, missing observations, irregular sampling, and uncertainty in
environmental inputs.

Second, the performance of the SINDy framework depends on the availability of
sufficiently informative temporal data. Typical multi-environment trial data,
which may contain only a small number of measurements per season, can make
derivative estimation and model identification difficult. Our sensitivity
analyses illustrate the effects of observation noise, sampling density, and
derivative-estimation settings, but are only a first step toward an extensive
robustness assessment. Increasing availability of high-throughput phenotyping
data may help address this limitation, together with appropriate smoothing and
derivative-estimation methods. Weak- or integral-form SINDy
methods~\citep{messenger2021weak,wei2022sparse} may provide useful alternatives
for sparse or noisy field observations, but were not evaluated here.
Furthermore, stochastic extensions of
SINDy~\citep{boninsegna2018sparse,fung2024rapid} seem promising for modelling
the inherent stochasticity of crop growth and environmental variation.

Third, sparse identification does not guarantee uniqueness of the resulting
equations. Correlated environmental variables and polynomial candidate
functions allow different combinations of terms to describe similar fitted
growth-rate responses. In our experiments, support stability varied across
library compositions and polynomial degrees, while cultivar-specific
coefficient differences and the resulting growth-rate functions remained more
stable at their respective levels of description. Individual polynomial terms
should therefore not be interpreted as uniquely recovered physiological
mechanisms. Because the identified polynomial rate functions are unconstrained
outside the range of crop states and environmental conditions represented in
the training data, extrapolation may produce physiologically implausible
behaviour, including negative states or non-monotone storage-organ
accumulation. In future work, Bayesian approaches that can quantify uncertainty
in both model structure and coefficients may help address this limitation.

Finally, the strong overall yield prediction partly reflects the dominant
year-to-year environmental variation in the dataset. The year-centred,
within-year ranking, and G$\times$E analyses provide a more demanding
assessment of cultivar-specific responses. The measured-LAI formulation
performed particularly well on these metrics, but all analyses were conducted
on the same five cultivars used during model identification. Prediction for
previously unseen genotypes and integration with genomic information remain
important topics for future work.

In summary, SINDy-based identification of reduced, explicit ODE models provides
a promising approach for learning crop-growth dynamics from longitudinal data.
In this controlled \textit{in silico} setting, the models reproduced growth
trajectories and yield variation across held-out years with predictive accuracy
comparable to dense polynomial and nonlinear dynamic alternatives. The
identified GSCs also retained reproducible cultivar-specific differences across
calibration environments, supporting their interpretation as compact
descriptors of cultivar-specific growth dynamics. Although the exact polynomial
representation was not always uniquely determined, alternative representations
could produce similar fitted growth-rate responses. Future work should test the
framework on real phenotyping data and investigate how GSCs and derived
functional characteristics can be linked to genomic information for prediction
of previously unseen genotypes.

\section{Conclusion}\label{sec:conclusion}
In this paper, we applied the Sparse Identification of Nonlinear Dynamics
(SINDy) framework to identify reduced ODE models of crop-growth dynamics
directly from longitudinal data. Using an \textit{in silico} dataset generated
with WOFOST, we showed that the identified models can reproduce storage-organ
growth trajectories and final yield in held-out years while retaining an
explicit, linear-in-parameters representation. Although the exact polynomial
representation was not uniquely determined by the available data, likely due to
the ill-posed nature of the identification problem and strong correlations
among environmental variables and candidate functions, the GSCs retained clear
cultivar-specific differences across calibration-environment resampling, while
the resulting growth-rate functions also preserved distinct
environmental-response patterns among cultivars. Future work will focus on
validation with real phenotyping data, uncertainty-aware statistical
extensions, and integration with genomic prediction.

\section{Code}\label{sec:code}
All code used to generate the results in this paper is available at
\url{https://github.com/CropXR/sindy-crop-growth}. Instructions for reproducing
the results and for applying the method to new datasets, including real-world
phenotyping data, are provided in the README file.

\section*{Competing interests}
No competing interests are declared.

\section*{Author contributions}
C.V. contributed to study design and methodology, implemented and performed
numerical experiments, conducted the analyses, and wrote the initial draft of
the manuscript. N.B. supervised the research, contributed to the study design
and methodology, interpretation of the results, and helped write the
manuscript. F.v.E. contributed to the interpretation of results through
discussions and provided feedback on the manuscript. A.H. contributed through
discussions and manuscript review. All authors approved the final manuscript.

\section*{Data availability}
The data and code required to reproduce the analyses presented in this study
are publicly available at \url{https://github.com/CropXR/sindy-crop-growth}.
The code repository contains the files to regenerate the \textit{in silico}
dataset, and the scripts required to reproduce the analyses, figures, and
tables in the manuscript.

\section*{Acknowledgments}
The authors thank the anonymous reviewers for their valuable suggestions. The
authors acknowledge the use of computational resources of the DelftBlue
supercomputer, provided by the Delft High Performance Computing
Centre~\citep{DHPC2024}. This publication is part of the research program of
the CropXR Institute financed by the National Growth Fund (NGF) of the
Netherlands and the Dutch Research Council (NWO) under the PlantXR programme
(project number KICH3.LTP.20.005).

The authors acknowledge the use of ChatGPT (by OpenAI) for language editing.
The tool was not used to generate scientific content or intellectual output.
All interpretations, analyses, and conclusions are solely those of the authors,
who take full responsibility for this manuscript.

\bibliographystyle{plainnat}
\bibliography{ref}

\end{document}